\documentclass{report}

\title{Up to 58 Tets/Hex to untangle Hex meshes}
\author{Luca Schaller}
\date{\today}

\usepackage{amsmath}
\usepackage{amssymb}
\DeclareMathOperator*{\argmin}{arg\,min}
\usepackage{graphicx}
\graphicspath{ {./graphics/} }
\usepackage{wrapfig}
\usepackage{algorithm}
\usepackage{algpseudocode}
\usepackage{float}
\usepackage{hyperref}
\usepackage{xcolor}

\hypersetup{
    colorlinks,
    linkcolor={red!40!black},
    citecolor={blue!70!black},
}

\begin{document}

\begin{titlepage}
    \newcommand{\HRule}{\rule{\linewidth}{0.5mm}}
    
    \center

    
    \includegraphics[width=0.3\textwidth]{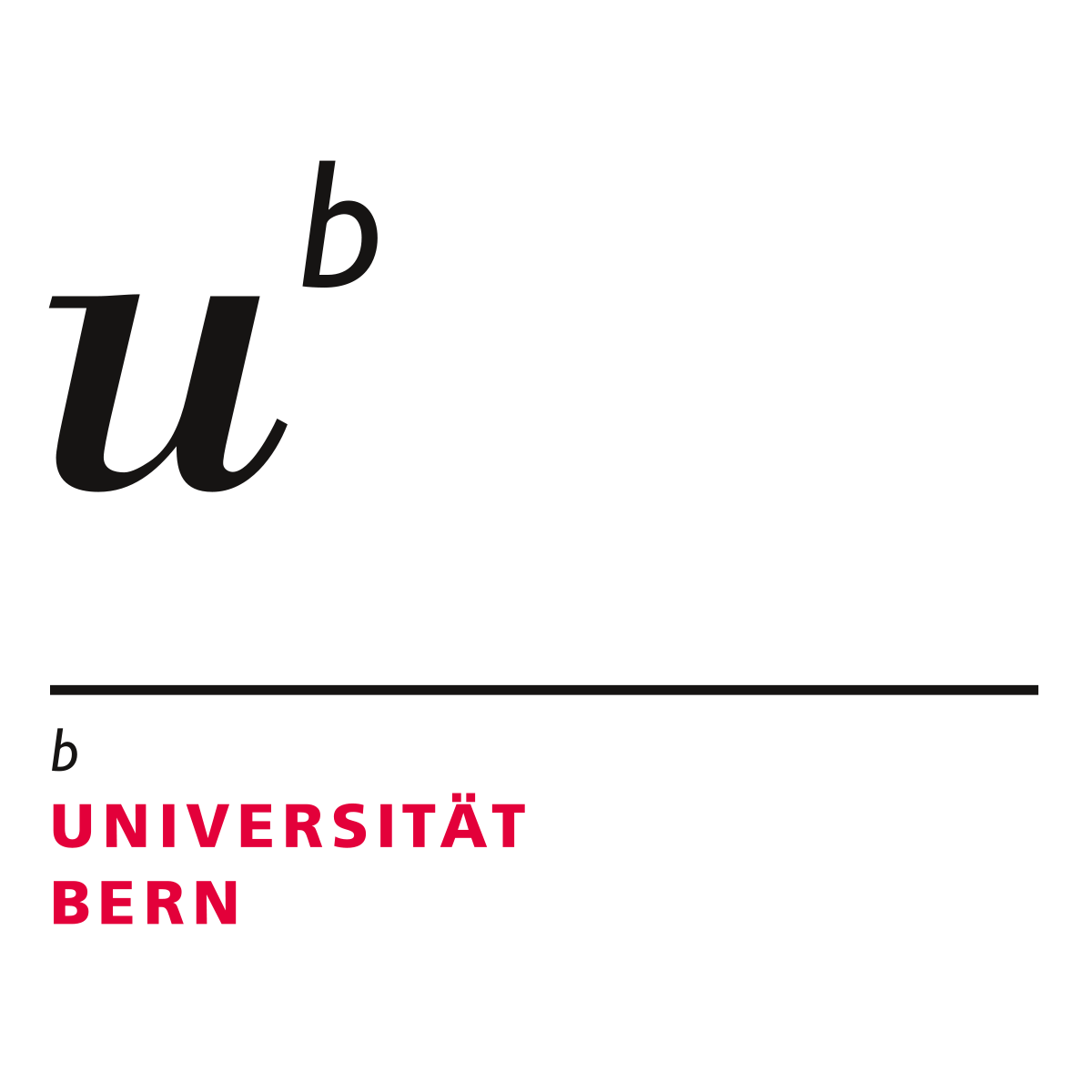}\\[1cm]
    
    
    \textsc{\LARGE University of Bern }\\[1.5cm]
    
    \textsc{\Large Bachelor Thesis}\\[0.5cm]

    \textsc{\large Computer Graphics Group}\\[0.5cm]
    
    
    \HRule\\[0.4cm]
    
    {\huge\bfseries Up to 58 Tets/Hex to untangle Hex meshes}\\[0.2cm]
    
    \HRule\\[1.5cm]
    
    
    \begin{minipage}{0.4\textwidth}
        \begin{flushleft}
            \large
            \textit{Author}\\
            Luca \textsc{Schaller}
        \end{flushleft}
    \end{minipage}
    ~
    \begin{minipage}{0.5\textwidth}
        \begin{flushright}
            \large
            \textit{Supervisors}\\
            Dr. Pierre-Alexandre \textsc{Beaufort}\\
            Prof. Dr. David \textsc{Bommes}
        \end{flushright}
    \end{minipage}
    
    
    \vfill\vfill\vfill
    
    {\large\today}
    
    \vfill
    
\end{titlepage}

\pagenumbering{roman}

\section*{\huge Abstract}

The request for high-quality solutions continually grows in a world where more and more tasks are executed through computers.  This also counts for fields such as engineering, computer graphics, etc., which use meshes to solve their problems. A mesh is a combination of some elementary elements, for which hexahedral elements are a good choice thanks to their superior numerical features. The solutions reached using these meshes depend on the quality of the elements making up the mesh. The problem is that these individual elements can take on a shape which prevents accurate computations. Such elements are considered to be invalid. To allow users to get accurate results, the shape of these elements must therefore be changed to be considered valid.

In this work, we combine the results of two papers to scan a mesh, identify possible invalid elements and then change the shape of these elements to make them valid. With this combination, we end up with a working algorithm. But there is room for improvement, which is why we introduce multiple improvements to speed up the algorithm as well as make it more robust. We then test our algorithm and compare it to another approach. This work, therefore, introduces a new efficient and robust approach to untangle invalid meshes.

\newpage

\tableofcontents
\newpage

\pagenumbering{arabic}

\chapter{Introduction}
Computer graphics, engineering and other areas rely on the description of a domain of some objects. These descriptions are, for example, used to run some finite element simulations on these objects. This way, an engineer could run a stability simulation on such a description instead of having to build a physical prototype. Depending on the application, it is sufficient to describe this object's surface, while the whole volume needs to be described at other times. Regardless of the application, the implementation is usually done using a mesh. A mesh is a combination of some space-filling primitive shapes. For example, a set of triangles conformally placed next to each other to describe the surface of some object. This thesis focuses on the 3D case. This means the whole volume of these objects is described. Therefore, some volumetric primitive shape has to be used. Due to their superior numerical features, we work with hexahedral elements.

These hexahedral elements are defined as a trilinear mapping of a unit cube into some physical space. A trilinear map is a function of three variables such that this function is linear separately in each of these three variables. In this case, the mapping is trilinear because each corner, or better-called vertex, of the unit cube, has three coordinates, and all three of these coordinates are linearly mapped into this physical space. A Jacobian matrix describes the mapping. From now on, this Jacobian matrix is simply called the Jacobian. Its definition follows in section \ref{sec:spaces}. Motivated by these superior numerical features, a lot of work has been put into algorithms which automatically generate hexahedral meshes. However, yielding a valid mesh is nontrivial. The validity of the whole mesh depends on each hexahedron being valid. If that is the case, the algorithm has achieved its goal.

The validity of a hexahedron depends on whether its trilinear mapping is injective or not. An injective function is also known as a one-to-one function. It is a function which maps distinct elements of its domain, all possible values for x, to distinct elements of its codomain, the resulting values $f(x)$. Therefore, in case $f(x_1) = f(x_2)$, it follows that $x_1$ and $x_2$ also have to be equal. So if the mapping is injective, the hexahedron takes on a valid form. The injectivity of this trilinear mapping can be checked by looking at the Jacobian determinant. The Jacobian determinant represents the volume of the mapped element. The problem is that the Jacobian of a hexahedron is not a constant but a polynomial function. Yet, this can be approximated by splitting the hexahedron into tetrahedra. For these tetrahedra, the Jacobian can then be assumed to be constant. These tetrahedra together then make up the volume of the hexahedra. If this volume is strictly positive everywhere, then the mapping is injective, and the element is valid.

Our contribution with this work is the combination of two papers to allow for hexahedral meshes to be checked for their validity and enable the optimisation of these meshes in case they are invalid. Therefore allowing such an invalid hexahedral mesh to be generated. Our algorithm can then do this finalising step of making the mesh valid. The validity check follows from the paper "robust and efficient validation of the linear hexahedral element" \cite{validation}, and the optimisation is based on the results of the paper "Foldover-Free Maps" \cite{foldover}. We also show several steps that can be taken to make the algorithm run faster or make it more robust.

\section{Related Works} \label{chap:related-work}
There are different approaches to untangle an invalid mesh. Three other untangling strategies are summarised in the three following sections. The paper discussed in section \ref{sec:sos} provides test cases also used in this thesis to compare results. This way, a direct comparison of the speed and limits of the algorithms can be presented.

\subsection{Untangling curvilinear Meshes}
\begin{figure}[H]
\centering
\includegraphics[scale=0.4]{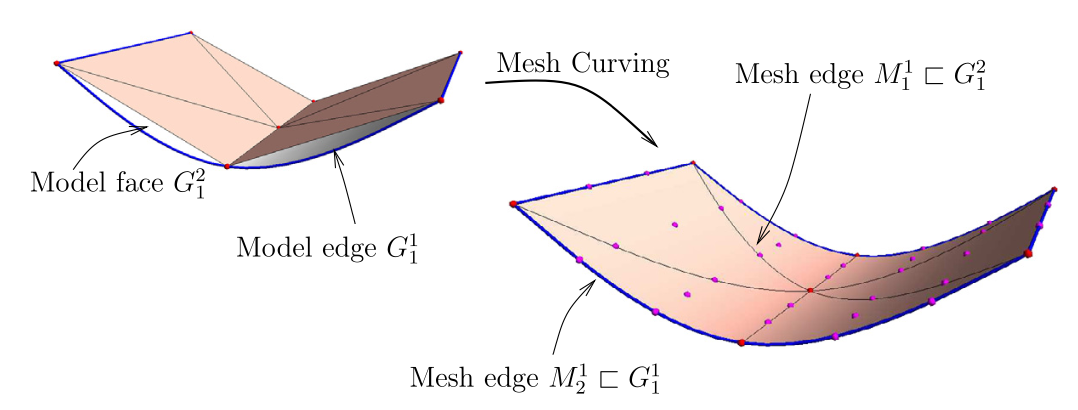}
\caption{the process of generating a high order mesh, Fig. 1 in \cite{curvlinear}}
\label{fig:curv-creation}
\end{figure}

\begin{figure}[H]
\centering
\includegraphics[scale=0.5]{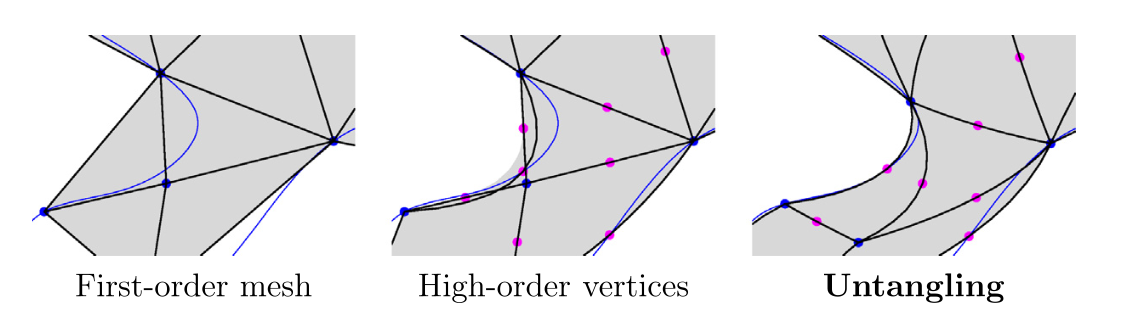}
\caption{the process of curving the mesh and then untangling, Fig. 2 in \cite{curvlinear}}
\label{fig:curv-untangling}
\end{figure}

The "Robust untangling of curvilinear Meshes" paper \cite{curvlinear} introduces an optimiser for curved meshes. Curving the mesh allows it to, for example, better fit a computer-aided design, CAD, model's geometry. Curving the mesh is done by defining high-order mappings between the reference domain and the physical shape. In order to do that, a high-order mesh has to be generated. They begin generating this high-order mesh by first generating a straight-sided mesh. Afterwards, high-order mesh points are added to the parts of the mesh where the original model is curved. This process is illustrated in Fig. \ref{fig:curv-creation}.
The position of these high-order mesh points can be chosen in different ways. For example, the distance between their position on the CAD model and their position on the mesh is minimised. The problem with this process of curving the straight-sided mesh is that elements might get tangled up. This means these elements are not valid anymore and need to be untangled. This process is illustrated in Fig. \ref{fig:curv-untangling}.
\noindent
The validity of these elements can be checked by finding some bounds: 
$$J_{min} = \min_{\xi} J \ \textrm{and} \ J_{max} = \max_{\xi} J$$
where $\xi$ represents the coordinates of the vertices in some reference space. These bounds can be computed by expressing the Jacobian as a Bézier polynomial. Due to the values of these bounds being contained within the range of these Bézier polynomial coefficients:
$$J(\boldsymbol \xi) = \sum_{i=1}^{N_q} B_i^{(q)}(\boldsymbol \xi)B_i$$
Where q is the order of the Bernstein polynomial $B_i$ of the Jacobian. This allows the calculation of bounds:
$$\min_{\xi} J(\xi) \geq \min_i B_i \ \textrm{and} \ \max_{\xi} J(\xi) \leq \max_i B_i$$
These bounds can be tightened with the "de Casteljau algorithm" \cite{deCasteljau}. Once the validity is checked, an objective function is presented with which it is possible to untangle curvilinear meshes. This objective function is split into two parts.
$$f = E + F$$
Where the first part, E, is responsible for penalising the movement of the vertices, as the mesh should stay as similar to the original as possible. The second part, F, represents a condition responsible for the mesh's elements to become valid. This condition is also implemented as a penalty function, which is achieved by using an underlying function:
$$F_{\epsilon}(x) = (x-1)^2 + \log \left ( \frac{x-\epsilon}{1-\epsilon} \right )^2$$
Therefore allowing this objective function to be minimised by the use of a moving-barrier method \cite{barrier-method}.

\subsection{SOS Relaxation} \label{sec:sos}
The "Hexahedral Mesh Repair via Sum-of-Squares Relaxation" paper \cite{sos} presents a method to validate and optimise a mesh through "SOS Relaxation", which stands for Sum-of-Squares Relaxation. This is a general technique for approximating solutions to problems whose objective and constraint functions have polynomials as their variables. Hence, this technique fits the analysis of hexahedral validity as hexahedral elements are polynomial elements. This SOS Relaxation can be used to check whether a hexahedral element is valid. This can be done by finding a sum-of-square decomposition of the variables where a sum-of-squares polynomial can be written as:
$$\det (\boldsymbol x) = \sum_{i=1}^n (q_i(\boldsymbol x))^2$$
where $\boldsymbol x = (x_1,...,x_n)$ and q is a real polynomial $q(x_1,...,x_n) \in \mathbb{R}[x_1,...,x_n]$. But this can also pinpoint the most tangled point. This is done by using moment relaxation \cite{moment-relax} to find the point $u^*$ where the determinant is minimised. Knowing this point allows them to untangle a hexahedron by moving its vertices as little as possible such that the hexahedron becomes injective at this point. Therefore ending up with this non-linear optimisation to untangle a hexahedron:
$$X^{k+1} = \argmin_X ||X-X^k||_F^2 \ s.t. \ \det (J(x(u^*))) \geq 0$$
where
$X = \begin{bmatrix}
x_{0,0,0} \\
\vdots \\
x_{1,1,1}
\end{bmatrix}$
represents a hexahedron with its 8 vertices.

With these two steps, an algorithm can be produced that checks a mesh for invalid elements and can solve these elements individually. But since the problem revolves around dealing with a mesh and not individual hexahedra, along with the fact that the optimised mesh should be as similar to the original mesh as possible, the algorithm is constrained. Therefore, the algorithm is again split into two steps. In the first step, individual hexes are repaired and followed by a second step which then brings these individually optimised hexes together in such a way that the mesh still looks similar. This balance between optimisation and keeping form is adjustable by some parameter $\rho$. Leaving them with an overall problem with the form:
$$\min_V ||V-V^0||_F^2 + \rho \sum_{\eta}||\Pi (H_{\eta}V)-H_{\eta}V||_F^2$$
Where $\eta$ represents the hexahedra, V are the vertex positions, $H_{\eta}$ are the vertices of hexahedron $\eta$, and $\Pi$ stands for the repair operator with the optimisation problem as defined above. Minimising this should untangle the mesh while keeping the movement of the vertices to a minimum.

\subsection{Edge-Cone Rectification}
\begin{figure}[h]
\centering
\includegraphics[scale=0.6]{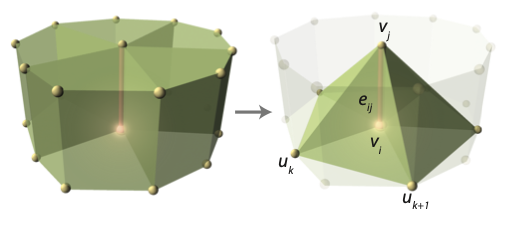}
\caption{edge-cone build from hexahedra, Fig. 2 b) from \cite{edge-cone}}{}
\label{fig:edge-cone_vertex}
\end{figure}

\begin{figure}[h]
\centering
\includegraphics[scale=0.8]{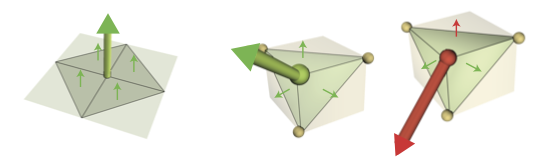}
\caption{edge-cone quality definition, Fig. 3 a) and b) from \cite{edge-cone}}
\label{fig:edge-cone_quality}
\end{figure}

In the "Practical Hex-Mesh Optimization via Edge-Cone Rectification" paper \cite{edge-cone}, another optimisation is presented to untangle hexahedral meshes. The optimisation is done by optimising the shape of the edge-cones. An edge-cone is a structure consisting of all tetrahedra considered to be corner tetrahedra in their hexahedron and follow the direction of a directed edge.
\begin{wrapfigure}{l}{0.3\textwidth}
\includegraphics[width=0.3\textwidth]{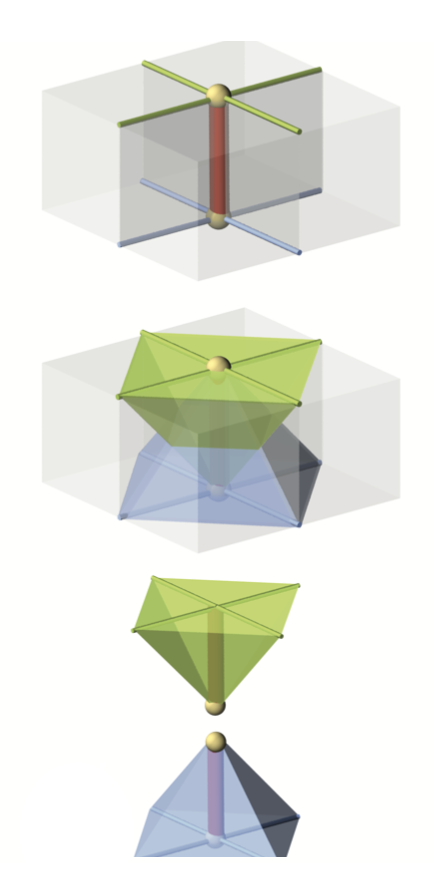}
\caption{edge-cones per directed edge, Fig. 1 d) from \cite{edge-cone}}
\label{fig:edge-cone-basics}
\end{wrapfigure}
Every edge in a mesh consists of two directed edges, one for each direction. This means for every edge there are two cones. 
For example, for a directed edge used by four hexahedra, four tetrahedra are building the cone, one corner tetrahedron from each hexahedron. This is illustrated in Fig. \ref{fig:edge-cone-basics}. When done for each directed edge, all corner tetrahedra of the mesh are included. The vertex from which the directed edge starts, $v_i$ in Fig. \ref{fig:edge-cone_vertex}, together with the two vertices $u_k$ and $u_{k+1}$, which are not part of the directed edge, form a plane when connected. Meaning if the cone consists of four hexahedra, there are four of those planes. The planes are illustrated in Fig. \ref{fig:edge-cone_quality}. For each plane, its normal vector is calculated. Edge-cone quality is then defined by the angle between the direction of the directed edge and each of the plane's normal vectors. All these directions are parallel in a perfect edge-cone, as illustrated on the left in Fig \ref{fig:edge-cone_quality}. An edge-cone becomes inverted when the angle between the direction of the directed edge and the direction of at least one plane's normal vector is greater than 90°. In Fig. \ref{fig:edge-cone_quality} if the edge cone direction is red, the edge-cone is invalid.

Therefore, an algorithm to optimise meshes can be built. A first try with an easier global formulation does not work, as the algorithm may terminate too early.
Therefore, they are using a two-step optimisation which consists of a local and a global step. During the local step, a target direction for the directed edge is computed. For the global step, the goal is to align the actual mesh edges with these target edge directions.
This is done by minimising an energy function consisting of four terms:
$$E = E_{Qcone} + E_{penalty} + E_{regularize} + E_{boundary}$$

The first and primary term is responsible for optimising the quality of an edge-cone:
$$E_{Qcone} = \sum_{e_{ij} \in E} \sum_{k=1}^{|T(e_{ij})|} (e_{ij} / L_{ij} - n_k)^2$$
Where $e_{ij}$ is a directed edge connecting vertices $i$ and $j$, $T(e_{ij})$ are the vertices which are directly connected to the vertex $i$ and belong to a hexahedron which uses the edge $e_{ij}$, $L_{ij}$ is the estimated length an edge will have in the target output and $n_k$ is the target direction calculated in the local step.

The penalty term keeps the new edge directions close enough to the target directions so that inversions do not occur. This is done by a penalty function instead of a hard constraint, allowing the algorithm to continue running even when these constraints cannot be met immediately. This way, premature termination can be avoided.

A regularisation factor is added in order to stabilise the algorithm in cases where the input mesh has a lot of invalid elements. Because there can be many different solutions for the local step. As a result, two halves of the same edge may have different target directions. Therefore, the edge directions should be generated more unified.

The boundary of a mesh should be preserved. But sometimes, some boundary vertices must move to untangle elements, so a boundary factor is also necessary. It balances the improvement of the mesh quality against boundary preservation.
$$E_{boundary} = \sum_{v \in S} \beta (\hat n \cdot V + \hat d)^2 + \sum_{v \in F} (\alpha (v-(\hat v + a \hat t ))^2 + a^2 ) + \sum_{v \in C} \alpha (v-\hat v )^2$$
Where a distinction between three types of vertices is made, there are regular surface vertice $v \in S$. Vertices which are part of a feature $v \in F$. And there are corner vertices which are part of multiple features $v \in C$. These three different types all have their individual energy function. $\hat{v}$ represents the reference position of each vertex, $\hat{n}$ is the input surface normal at that point, and $\hat{t}$ represents the feature tangent at $\hat{v}$. The two parameters $\alpha$ and $\beta$ are weighting parameters.

\newpage

\chapter{58 Tetrahedra/Hexahedron}
In our approach, the results of two main papers are used. Firstly, "robust and efficient validation of the linear hexahedral element" \cite{validation}, which shows a sufficient condition for the validity of a hexahedral element, through the use of the results from "Nondegeneracy tests for hexahedral cells" \cite{nondegeneracy}. Secondly, "Foldover-Free Maps" \cite{foldover} provides formulas and a scheme to untangle tetrahedral meshes. We, therefore, look at our hexahedra as combinations of tetrahedra to make use of their result. When combining these two steps, a method can be developed which looks at a hexahedral mesh, determines its invalid elements and optimises these invalid hexahedra to take on a valid form. As the mesh should keep its general shape as it is, the boundary vertices of the mesh are fixed. This means they are not allowed to move in order to untangle the mesh, while the inner vertices are free to move. The boundary relaxation is addressed in section \ref{sec:movable-bnd}.

\section{Validation of Hexahedra} \label{sec:validity}
When looping over a mesh to uncover invalid elements, a condition which can tell whether a hexahedral element is valid or not is needed. There are multiple possible conditions to check this. Here, we address one pretty simple condition using tetrahedra.

When taking the 8 vertices that make up a hexahedron and always combining 4 of them into a tetrahedron, 70 possible tetrahedra can be built. This follows from the binomial coefficient $\binom{8}{4} = 70$. But out of these 70 tetrahedra, only 58 have a non-zero volume. And since a tetrahedron with a zero volume does not reveal any information about the form of a hexahedron, only 58 tetrahedra are needed. It is possible to compute these 58 tetrahedra with a brute-force approach. Starting with a unit cube, four nested loops are then used. The outermost loop goes from 0 to 4, the second from 1 to 5 and so on, storing these combinations. Their volume can be computed with the determinant of their Jacobian, whose definition follows in section \ref{sec:spaces}. If the volume is zero, the tetrahedron is discarded, and otherwise, it is stored. Additionally, the volume might also be negative, as the order of vertices in the tetrahedra is chosen arbitrarily, depending on how the tetrahedra creation algorithm works. But that can easily be fixed by just swapping two vertices. 
\begin{wrapfigure}{l}{0.2\textwidth}
\includegraphics[width=0.2\textwidth]{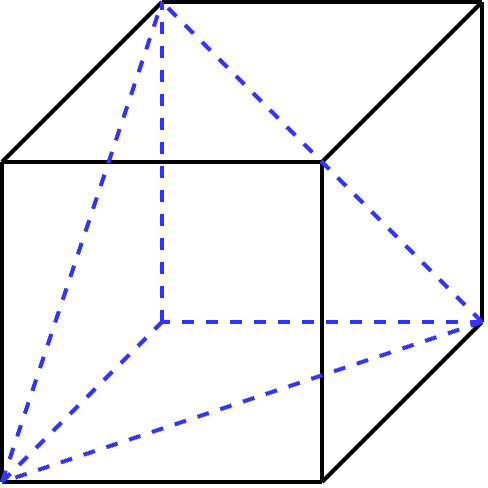}
\caption{one of the 8 possible corner tetrahedra}
\label{fig:corner-tet}
\end{wrapfigure}
Of these 58 tetrahedra, 8 of them are considered corner tetrahedra. A corner tetrahedron consists of a vertex and the three vertices with which this vertex builds an edge of the cube. As illustrated on the left, the dashed blue lines represent one of the possible 8 corner tetrahedra in a hexahedron. It is known that for a hexahedron to be valid, it is a necessary condition that these 8 corner tetrahedra all have a positive volume. Necessary here does not mean a hexahedral element is valid, but it might be valid. While if not all 8 have a positive volume, we know that the hexahedron is invalid.

In order to get a sufficient condition on whether a hexahedral element is valid or not, we turn to "Nondegeneracy tests for hexahedral cells" \cite{nondegeneracy}. According to this paper, the smallest amount of tetrahedra with a positive determinant necessary to end up with a sufficient condition for the validity of a hexahedral element is all 58 tetrahedra. Sufficient here means that it is certainly valid. While, if not all 50, as the 8 corner tetrahedra are necessary, have a positive volume, it is unknown whether the hexahedron is valid or not. Sadly this sufficient condition also classifies many actually valid elements as invalid because the condition is too strong. This is illustrated in \cite{validation} Table 4, the columns labelled "Ours" and "Test5". According to \cite{validation}, this sufficient condition can classify up to 80\% of actually valid hexahedra as invalid. As they created an algorithm using a different validity check which is also sufficient but does not classify any actual valid elements as invalid. This algorithm is looked at in section \ref{sec:new-validity}. As the difference in the number of hexahedra classified as invalid is relevant.

\section{Foldover-Free Maps}
This section is taken from "Foldover-Free Maps" paper \cite{foldover}. Their formulas are written in a generic way for 2D and 3D, represented by a variable $d$. We adapt this part to be specific to our case, which is the 3D case, d = 3.

\subsection{Three Linear Map Spaces} \label{sec:spaces}
The problem we are trying to solve presents hexahedra which have some shape in a mesh. One hexahedron is made up of 58 tetrahedra. All of these 58 tetrahedra have their own shape, which is called their physical shape. When looking at the same tetrahedron in two different hexahedra, they most likely do not have the same physical shape. Next to their physical shape, all 58 tetrahedra also have a target shape. This respective target shape would be the goal for all 58 tetrahedra to reach through the optimisation process. In our case, there are 58 target shapes in total, one for each tetrahedron. So tetrahedra from different hexahedra refer to the same target shape. Meaning that when looking at the same tetrahedron in two different hexahedra, they have the same target shape. The target shape corresponds to the topological tetrahedron in the unit cube.

\begin{figure}[H]
\centering
\includegraphics[scale=0.6]{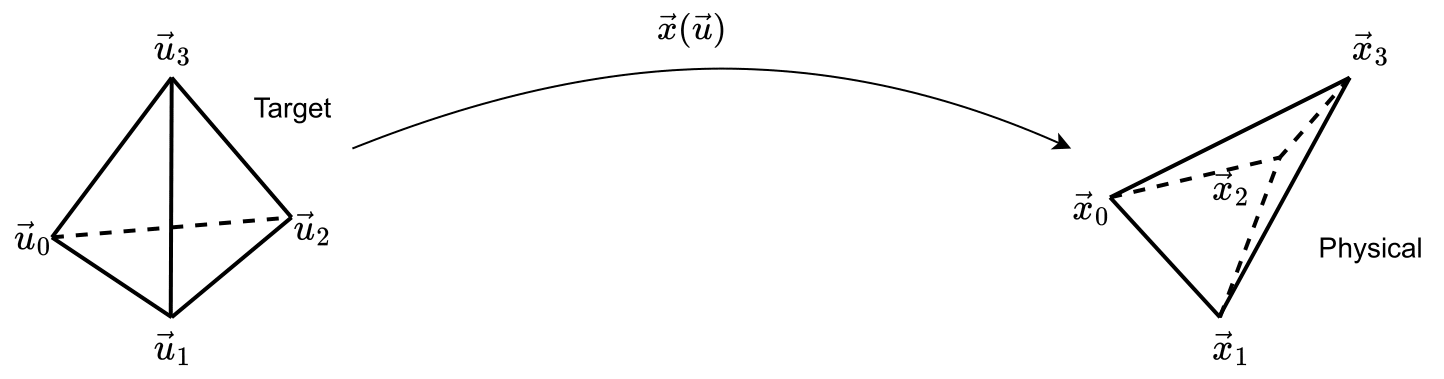}
\caption{the mapping from the target to the physical space}
\label{fig:mapping}
\end{figure}

As illustrated in Fig. \ref{fig:mapping}, there exists a mapping $\vec x$ between the two shapes. In order to optimise this mapping, its Jacobian is needed. The Jacobian stands for the differential of our mapping. Therefore the Jacobian of mapping $\vec x$ is built:
$$J = \frac{\partial \vec x}{\partial \vec u} = 
\left ( \frac{\partial \vec x}{\partial u_1} ... \frac{\partial \vec x}{\partial u_3} \right ) =
\begin{pmatrix}
\frac{\partial x_1}{\partial u_1} & ... & \frac{\partial x_1}{\partial u_3}\\
: &  & :\\
\frac{\partial x_3}{\partial u_1} & ... & \frac{\partial x_3}{\partial u_3}
\end{pmatrix}$$

\begin{figure}[H]
\includegraphics[scale=0.55]{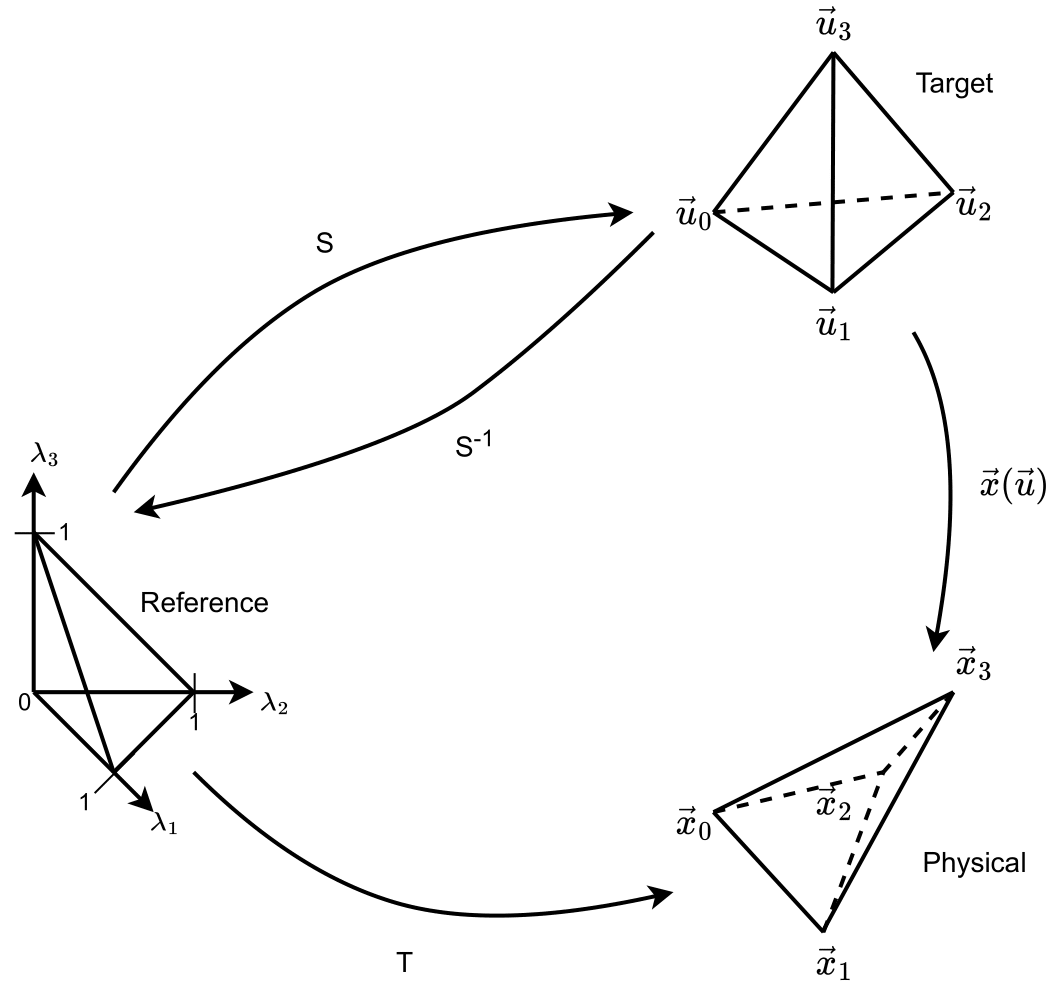}
\centering
\caption{the 3 different spaces which are used to optimise a tetrahedron}
\label{fig:spaces}
\end{figure}

Due to this mapping $\vec x (\vec u)$ in Fig. \ref{fig:mapping} from the target to the physical space being uneasy to obtain directly, another third space is introduced. It is called the reference space. This space expresses the tetrahedra in barycentric coordinates as illustrated in Fig. \ref{fig:spaces}.

\paragraph{Barycentric Coordinates}
The barycentric coordinate system for tetrahedra is well suited for our computational needs. In our 3D case, the barycentric coordinate system is described by 4 coordinates called $\lambda_0, \lambda_1, \lambda_2$ and $\lambda_3$, to represent every point on the tetrahedron with respect to its 4 vertices. To define any point $\boldsymbol r = (x,y,z)$ in these coordinates, the coordinates of the corners of the tetrahedron have to be taken and multiplied with their respective lambda. 
$$\boldsymbol r = \lambda_0 \boldsymbol x_0 + \lambda_1 \boldsymbol x_1 + \lambda_2 \boldsymbol x_2 + \lambda_3 \boldsymbol x_3$$
There are two conditions these 4 coordinates need to fullfil:
$$\lambda_i \geq 0 \ \textrm{for} \ i = 0,1,2,3 \ \textrm{and} \ \lambda_0 + \lambda_1 + \lambda_2 + \lambda_3 = 1$$
The equation above can be simplified by reformulating the second condition:
$$\lambda_0 = 1 - \lambda_1 - \lambda_2 - \lambda_3$$
Which then leaves us with:
$$\boldsymbol r = (1 - \lambda_1 - \lambda_2 - \lambda_3) \boldsymbol x_0 + \lambda_1 \boldsymbol x_1 + \lambda_2 \boldsymbol x_2+ \lambda_3 \boldsymbol x_3$$

Now that there is a representation in barycentric coordinates at hand, the Jacobian of the mapping between these two coordinate systems can be calculated. Using its definition from above combined with the partial derivatives, it follows for mapping T:
$$ J(T) =
\begin{pmatrix}
x_{1} - x_{0} & x_{2} - x_{0} & x_{3} - x_{0}\\
y_{1} - y_{0} & y_{2} - y_{0} & y_{3} - y_{0}\\
z_{1} - z_{0} & z_{2} - z_{0} & z_{3} - z_{0}
\end{pmatrix}$$
Where x, y and z are the three coordinate components of the vertices.

With this newly introduced space, the mapping $\vec x (\vec u)$ can be described and found much easier. As the goal is to go from the target to the physical space, the inverse of the mapping S has to be taken and connected with the mapping T, as shown in Fig. \ref{fig:spaces}. Therefore:
\begin{equation} \label{form:mapping} \vec x (\vec u) = T\cdot S^{-1} \end{equation}
The mappings S and T are described by their Jacobian.

\subsection{Energy} \label{sec:energy}
In order to optimise the shape of a tetrahedron and, therefore, the shape of its respective hexahedron, an objective function has to be defined. This is done using the energy function of "Foldover-Free Maps" \cite{foldover}. This paper supplies formulas to calculate the energy of a mesh with invalid simplicies and presents an optimisation scheme to minimise this energy. This results in a mesh where all physical elements tend to be as close as possible to their target shape, up to some margin.

Defining the map $\vec x: \Omega_{target} \subset \mathbb{R}^3 \to \mathbb{R}^3$ and starting with the following variational problem:
\begin{equation} \label{form:variational_primitive} \argmin_{\vec x} \int_{\Omega_{target}} (f(J) + \lambda g(J)) du \end{equation} 
with J being the Jacobian of the mapping $\vec x$. Recall from \eqref{form:mapping}.

\begin{equation} 
\label{form:f_primitive}
f(J) :=
\begin{cases}
\frac{tr J^TJ}{(\det J)^{\frac{2}{3}}}, &\det J > 0\\
+ \infty, &\det J \leq 0
\end{cases}
\end{equation}

\begin{equation} 
\label{form:g_primitive}
g(J) := 
\begin{cases}
\det J + \frac{1}{\det J}, &\det J > 0\\
+ \infty, &\det J \leq 0
\end{cases}
\end{equation}
\noindent
Where $f(J)$ is minimised when the shapes of the physical tetrahedron have the same angles as in the target tetrahedron. At the same time, $g(J)$ is minimised when the volume of the physical tetrahedron matches the target volume. These two functions pursue possibly conflicting goals and are balanced by the $\lambda$ parameter as defined in \eqref{form:variational_primitive}.

These formulas provide a way to improve a valid simplicial mesh. But that is not the point since our goal is to have a tangled mesh and untangle it. The problem is that in a tangled mesh, the tangled tetrahedra have a Jacobian determinant which is negative or even zero. Currently, these cases are respected by just assigning an infinite amount of energy to such an element, but this does not help in any way to solve the problem at hand.

Therefore a modification to the energy formulation needs to be introduced. This modification is realised by introducing a regularisation function $\chi$ for a positive value of $\epsilon$:
\begin{equation} \label{form:chi} \chi (\det J, \epsilon) := \frac{\det J + \sqrt{\epsilon^2 + \det^2 J}}{2}, \ \epsilon \geq 0 \end{equation} 
With this modification \eqref{form:f_primitive} and \eqref{form:g_primitive} change to be:
\begin{equation} \label{form:energy-parts-final} f_{\epsilon}(J) := \frac{tr J^TJ}{(\chi(\det J, \epsilon))^{\frac{2}{3}}}, \
g_{\epsilon}(J) := \frac{\det^2 J + 1}{\chi(\det J, \epsilon)} \end{equation}
Note that $f_{\epsilon = 0}(J) = f(J)$ and $g_{\epsilon = 0}(J) = g(J)$. This modification affects the whole variational problem \eqref{form:variational_primitive} and it, therefore, changes to be:
\begin{equation} \label{form:integral} \lim_{\epsilon \to 0^+} \argmin_{\vec x} \int_{\Omega_{target}} (f_{\epsilon}(J) + \lambda g_{\epsilon}(J)) dx \end{equation} 
While it is possible to work with an integral, the variational problem can be simplified by discretising it. This is possible since the map $\vec x$ is piecewise-affine with the Jacobian being piecewise constant. Therefore:
\begin{equation} \label{form:variational} \lim_{\epsilon \to 0^+} \argmin_X F(X, \epsilon) \end{equation}
with 
\begin{equation} \label{form:energy-final} F(X, \epsilon) := \sum^{\#T}_{t=1} (f_{\epsilon}(J_t) + \lambda g_{\epsilon}(J_t)) vol(T_t) \end{equation}
being the discretised form of the integral \eqref{form:integral}.

Where $X := (\vec x^T_1 ... \vec x^T_{\# V})^T$ is the vector containing all the variables, representing the whole mesh. Consequently, $\# V$ stands for the number of vertices in the mesh. $\# T$ stands for the number of tetrahedra in the mesh, $J_t$ is the Jacobian for tetrahedron $t$ of the mapping $\vec x$ and $vol(T_t)$ is the volume of the target tetrahedron $t$.

With this formulation untangling a mesh is possible since tangled elements have a bounded energy. The Implementation follows further down.

\subsection{Epsilon $\epsilon$} \label{sec:epsilon}
By minimising the energy of our mesh X iteratively, an appropriate update for $\epsilon$ at each step is necessary. The value of $\epsilon$ needs to be adapted at every optimisation step. 

As an iterative descent method is used to solve problem \eqref{form:variational}, one of the two following conditions has to be fulfilled at every step for some $0 < \sigma < 1$:
\begin{equation} \label{form:descent-condition} F(X^{k+1}, \epsilon^k) \leq (1-\sigma)F(X^k, \epsilon^k) \end{equation}
or
\begin{equation} \label{form:minimality-comdition} \min_X F(X, \epsilon^k) > (1-\sigma)F(X^k, \epsilon^k) \end{equation}
Where condition \eqref{form:descent-condition} is a descent condition which has to hold for the most part. Condition \eqref{form:minimality-comdition} is for the case when $X^k$ is very close to the global minimum $F(X, \epsilon^k)$ which prevents \eqref{form:descent-condition} to be satisfied. Notice the change in the inequality direction. Looking at \eqref{form:descent-condition} and assuming it holds at iteration k, the following inequality needs to be fulfilled in order to show the non-increasing nature of the function:
$$(1-\sigma)F(X^{k+1}, \epsilon^{k+1}) \leq F(X^k, \epsilon^k)$$
It can be shown that the following update rule for $\epsilon^{k+1}$:
\begin{equation} \label{form:epsilon-update} \epsilon^{k+1} = 2 \sqrt{\mu^k (\mu^k - D^{k+1}_{min})} \end{equation}
where 
\begin{equation} \label{form:mu-update} \mu^k := (1 - \sigma^k) \chi(D^{k+1}_{min}, \epsilon^k) \end{equation}
fulfils this condition. Additionally $D^{k+1}_{min} := \min_{t \in 1 ... \# T} \det J^{k+1}_t$ represents the minimum determinant on the whole mesh at iteration k+1.
Together with \eqref{form:descent-condition}, this update rule shows the non-increasing nature of the function. The same update rule is also viable for \eqref{form:minimality-comdition}. This just leaves $\sigma^k$ undefined. It stands for the descent coefficient, but as its global value $\sigma$ is not known at the beginning, it is calculated locally at every step: 
$$\sigma^k := 1 - \frac{F(X^{k+1}, \epsilon^k)}{F(X^k, \epsilon^k)}$$
As the update rules \eqref{form:epsilon-update} and \eqref{form:mu-update} are only allowed for cases where $\sigma^k \geq \sigma$. It has to be ensured that the validity of \eqref{form:minimality-comdition} is fulfilled by $\sigma$. Because that allows the assignment $\sigma^k = \sigma$. \cite{foldover} suggests $\sigma = \frac{1}{10}.$

The update process, therefore, starts with an $\epsilon^0 = 1$ and updates it using the following formula:
\begin{equation}
\label{form:epsilon-update-final}
\epsilon^k :=
\begin{cases}
2 \sqrt{\mu^k (\mu^k - D^{k+1}_{min})}, &D^{k+1}_{min} < \mu^k\\
0, & D^{k+1}_{min} \geq \mu^k
\end{cases}
\end{equation}
 with the definitions from this section and $\sigma^k = \max \left ( \frac{1}{10}, 1 - \frac{F(X^{k+1}, \epsilon^k)}{F(X^k, \epsilon^k)} \right )$.

\section{Global Algorithm} \label{sec:algorithm}
Now that all the parts are together, a simple implementation can be realised. The pseudocode in Algorithm \ref{alg:main} describes the procedure.

\begin{algorithm}[H]
\caption{58 Tetrahedra}\label{alg:main}
\textbf{Input}: X, List of Unit Tetrahedra\\
\textbf{Output}: X
\begin{algorithmic}
\State $\textrm{validity} \gets \textrm{check mesh validity}(X, Tetrahedra)$
\If {not validity}
	\State $\epsilon^0 \gets 1$
	\While{$\det\nolimits_{\rm min} \leq 0$}
		\State $F_{prev} \gets \textrm{energy}(X^k, \epsilon^k, Tetrahedra)$ (Formula \eqref{form:energy-final})
		\State $X^{k+1} \gets X^k + \Delta X$
		\State $F \gets \textrm{energy}(X^{k+1}, \epsilon^k, Tetrahedra)$ (Formula \eqref{form:energy-final})
		\State $\epsilon^{k+1} \gets \textrm{update epsilon}(F_{prev}, F, \epsilon^k)$ (Formula \eqref{form:epsilon-update-final})
	\EndWhile
\EndIf
\end{algorithmic}
\end{algorithm}

A simple check of whether the mesh needs to be untangled starts the algorithm. The initial $\epsilon$ is set, and the main loop starts. Calculate the initial energy of the whole mesh by calculating the energy for every element, which means $\# hexahedra * 58$ energy calculations and summing them up. Now it is the optimiser's time to move around the mesh's vertices and minimise the energy. The final energy is calculated the same way the initial one was calculated. To finish off one iteration, $\epsilon$ is updated. This process is repeated until the minimum determinant, the smallest determinant of all $\#hexahedra * 58$ tetrahedra, is greater than zero, implying that the mesh is valid.

One main thing to note is that the step updating the mesh is the main step of the whole algorithm. It depends heavily on the optimiser chosen for the problem and is the leading time consumer of the algorithm. Another important note is that the boundary nodes are fixed, as mentioned at the beginning of this chapter.

Algorithm \ref{alg:main} needs a mesh to optimise in addition to a list of tetrahedra. This list should contain all 58 target shape tetrahedra, but storing the inverse Jacobian of these tetrahedra is also recommended. This way, they do not need to be calculated. The determinant of their Jacobian, their volume, can also be stored if wanted.

This is the basic implementation. In these following subsections, a few improvements are introduced regarding the algorithm's speed or the limit on what kind of meshes can be handled.

\subsection{Energy Function} \label{sec:energy-function}
Recalling section \ref{sec:energy} the energy is computed through the influence of two terms $f(J)$ and $g(J)$. Where $g(J)$ is responsible for trying to match the volume of a physical tetrahedron to the volume of its respective target tetrahedron. This factor is mainly in place due to the fact that the mesh would collapse to a single point during the optimisation without it. But since the boundary vertices are fixed in our method, this factor is no longer needed as the mesh cannot collapse to a single point. Removing its contribution to the energy makes it easier for the optimiser to find the minimum of the energy function as fewer factors influence it.

\subsection{Epsilon Update} \label{sec:epsilon-update}
The $\epsilon$ update presented in section \ref{sec:epsilon} always leads to a valid solution if one exists. But its update is sometimes relatively slow and could be sped up for the whole untangling process to converge faster. The guarantee of convergence \cite{foldover} provides is therefore lost. But since the faster convergence update is only done under certain conditions and the slower version is still used otherwise, this has not proven to be a problem for our test cases.

The impact $\epsilon$ has on the convergence can be seen in equation \eqref{form:energy-parts-final}. After section \eqref{sec:energy-function} only $f_{\epsilon}(J)$ is considered. Looking at \eqref{form:energy-parts-final}, the numerator is responsible for the optimised element to have angles which are similar to its target shape. While the denominator, equation \eqref{form:chi}, represents the volume and scales the angles accordingly. When $\epsilon$ is set to 1 in the beginning, it mostly eliminates the impact the denominator has on elements with a determinant $\leq 0$. Therefore it is just the angles of these elements which want to be improved in the beginning, as this would lead to valid elements. For elements with a determinant $> 0$, both terms are relevant. And due to the elements with determinants closest to $\epsilon$ being pushed to move the most, these already valid elements are mainly changed. This possibly leads to higher numbers of invalid hexahedra than in the beginning. But by doing this, the denominator value of these elements with a determinant $> 0$ can be lowered. Allowing space to be created for the elements with a determinant $\leq 0$ to improve their angles and therefore untangle themselves. But as stated, the elements with determinants similar to $\epsilon$ are pushed to move the most. The problem is, in general, the invalid elements have significantly smaller determinants than 1. This means $\epsilon$ needs to be decreased. Currently the $\epsilon$ parameter is always updated according to equation \eqref{form:epsilon-update-final}. Since the energy is usually not decreased significantly in one iteration, the $\epsilon$ update only lowers $\epsilon$ by $10\%$ of its previous value most of the time. This results in a very slow convergence towards 0 for $\epsilon$. And as most determinants of invalid elements are quite small, it takes a long time until they are pushed to move and untangle. This is why the epsilon update has room for improvement. That is where this new $\epsilon$ update comes into play. As the security of a convergence wants to be kept as well as possible, this faster epsilon update is only applied under certain conditions. If the three conditions:
\begin{equation}\label{eq:epsilon-conditions}
\begin{aligned}
\textrm{(i)} &\ \det\nolimits_{\rm min} < 0 \\
\textrm{(ii)} &\ \epsilon > |\det\nolimits_{\rm min}| \\
\textrm{(iii)} &\ (\det\nolimits_{\rm min}^{k+1}-\det\nolimits_{\rm min}^k) / |\det\nolimits_{\rm min}^k| > 10^{-4}
\end{aligned}
\end{equation}
are all fulfilled, $\epsilon$ can be updated in this matter:
$$\epsilon = - 0.2 \cdot \det\nolimits_{\rm min}$$
Where $\det\nolimits_{\rm min}$ is always the minimum determinant on the whole mesh.

\subsection{Boundary Relaxation} \label{sec:movable-bnd}
As discussed at the beginning of this chapter, the boundary vertices need to be looked at. While, in general, it is a good idea to fix the boundary vertices and keep the mesh shape the same, it might prove to be a problem for some meshes where the surface mesh is already invalid, with the surface mesh being the quadrangle mesh described by the boundary vertices. These quadrangles are all a face of some hexahedron in the actual mesh. If one of these quadrangles is invalid, the mesh cannot be untangled without moving some boundary vertex.

Since the goal remains to keep the general shape of the mesh as similar to the original as possible, two main steps are needed to reach this goal while also reaching the goal of untangling the mesh. Firstly as few boundary vertices as possible are allowed to move. The most straightforward idea here is only to allow boundary vertices which are part of an invalid hexahedron, to move. Secondly, a penalty term has to be introduced for the movement of the boundary vertices. That way, they are not pushed further than really necessary in order to untangle the mesh. This penalty term is added to the total energy of the mesh. It is computed by:
$$F_{penalty} = (pos_{current} - pos_{start})^2 * factor$$
Where $pos$ stands for the vertex position where it is currently or at the start. For the value of the factor, we suggest $10^6$. But this factor and its impact are discussed again in section \ref{sec:influence-penalty-factor}.

Looking at when the boundary vertices should be allowed to move, that can be brought into combination with the previous section \ref{sec:epsilon-update}. The boundary stays locked as long as the fast $\epsilon$ update is viable. Meaning that the conditions \eqref{eq:epsilon-conditions} are fulfilled. As soon as this is not the case anymore, the boundary vertices are allowed to move. But as they move, they inflict this penalty energy on the whole mesh energy.

\subsection{Validity Condition} \label{sec:new-validity}
As mentioned in section \ref{sec:validity}, there is an algorithm that can check the validity of a hexahedral element much more accurately than the sufficient condition used to check the validity so far. As this reduces the amount of invalid hexahedra which need to be optimised, the algorithm is sped up. The algorithm is described in detail in \cite{validation}.

The algorithm is based on the properties of Bézier functions. Such a Bézier function for a hexahedron is expressed as a tensor product of three Bernstein polynomials:
$$B_{ijk}^n(\xi, \eta, \zeta) = B_i^n(\xi)B_j^n(\eta)B_k^n(\zeta)$$
Where $\xi, \eta \ \textrm{and} \ \zeta$ are the coordinates in the reference space. Equivalent to the $\lambda_1, \lambda_2 \ \textrm{and} \ \lambda_3$ we use. The Bernstein polynomials are defined as:
$$B_k^n(t) = \binom{k}{n} t^k (1-t)^{n-k} \ , \ t \in [0,1], \ k = 0,...,n$$
The Bézier function is currently written for order n of the hexahedral polynomial space. But for hexahedra, only order 2 is needed as the Jacobian of a hexahedron is a triquadratic function. Therefore:
$$J(\boldsymbol\xi) = \sum_{i,j,k=0}^2 b_{ijk}B_{ijk}^2(\boldsymbol\xi)$$
Where the $b_{ijk}$ are the so-called control values. There are 27 such control values which follow from combinatorics $3^3$. These control values make it possible to bound the minimum Jacobian determinant form below and above. If the upper bound is negative, the hexahedron is classified as invalid. But if the lower bound is positive, it is a valid element. The case where the lower bound is negative, and the upper bound is positive does not allow a classification of the element. But this case only appears because the accuracy of these two bounds is not ideal after their first computation. But this accuracy can be increased through the use of the "de Casteljau algorithm" \cite{deCasteljau}. This algorithm subdivides the current Bézier function, allowing for a more accurate representation of the actual form of the hexahedral triquadratic function. Thanks to this more precise representation, the bounds automatically become more accurate. This subdivision can be done until the upper, and lower bounds take on values which allow a classification of the hexahedral element.

\subsection{Amount of Tetrahedra/Hexahedron} \label{sec:amount-elements}
The amount of elements which affect the mesh and need to be optimised strongly dominate how fast a solution can be found. That is why it is a good idea to reduce the number of elements. Section \ref{sec:validity} shows that 58 tetrahedra are a sufficient condition for the validity. Therefore it only seems logical to use those 58 tetrahedra for the optimisation. This leads to $\# hexahedra \cdot 58$ elements making up the whole mesh, leading to many elements that need to be optimised. But as mentioned before, 58 tetrahedra is a condition which is too strong. With those facts, it appears logical to decrease the amount of tetrahedra used in the optimisation. While not having a guarantee of being able to untangle a hexahedron with less than 58 tetrahedra, in most cases, fewer tetrahedra suffice. As the 8 corner tetrahedra form a necessary condition and sum up to the lowest amount of elements, this is the starting point. Only using the 8 corner tetrahedra decreases the number of optimisation elements by a factor of $\frac{58}{8} = 7.25$. There are then two ideas about how to deal with cases where 8 tetrahedra are not enough to untangle the mesh.

First, take a look at every hexahedron individually. If their $\det\nolimits_{\rm min} > 0$ but they are still invalid, increase the amount of tetrahedra used for this hexahedron. Because this indicates that this hexahedron cannot be untangled by only using the 8 corner tetrahedra but that more are needed. The easiest option here is to increase the number to 58 immediately. Even though different strategies could be interesting. This is picked up again in section \ref{sec:amount-tets}.

Second, as a general fallback, in case the untangling process gets stuck. Stuck here being defined by the following conditions being satisfied:
\begin{equation}\label{eq:elements-conditions}
\begin{aligned}
\textrm{(i)} &\ 10^2 \cdot \epsilon < |\det\nolimits_{\rm min}|\\
\textrm{(ii)} &\ \#inv^{k+1}-\#inv^k = 0 \\
\textrm{(iii)} &\ |(\det\nolimits_{\rm min}^{k+1}-\det\nolimits_{\rm min}^k) / \det\nolimits_{\rm min}^k| < 10^{-12}
\end{aligned}
\end{equation}
The amount of tetrahedra used for the untangling can be increased to 58 again. But this time, instead of only doing it for individual hexahedra, the increase is done for all hexahedra, as these conditions form a general fallback.

It is important to note that when incrementing the amount of tetrahedra used for individual hexahedra is done before the $\epsilon$ value is updated. This order relevance comes from the fact that it is not impossible to have cases where the mesh is still invalid even if $\det\nolimits_{\rm min} > 0$. Due to the 8 corner tetrahedra only forming a necessary condition. If done in the wrong order the $\epsilon$ value is be set to 0 according to \eqref{form:epsilon-update-final} because \eqref{form:chi} returns 0. Which in turn causes a division by 0 in the energy calculation. Therefore not allowing the algorithm to continue untangling the mesh. When first increasing the amount of tetrahedra used for the necessary hexahedra, the correct minimum determinant is used to calculate $\epsilon$, bypassing this pitfall. Also, when the amount of tetrahedra has to be increased for all hexahedra through the general fallback, the $\epsilon$ has to be recomputed immediately after the increment, as the minimum determinant might take on a different value.

\subsection{Blobs} \label{sec:blobs}
\begin{figure}[H]
\centering
\includegraphics[scale=0.5]{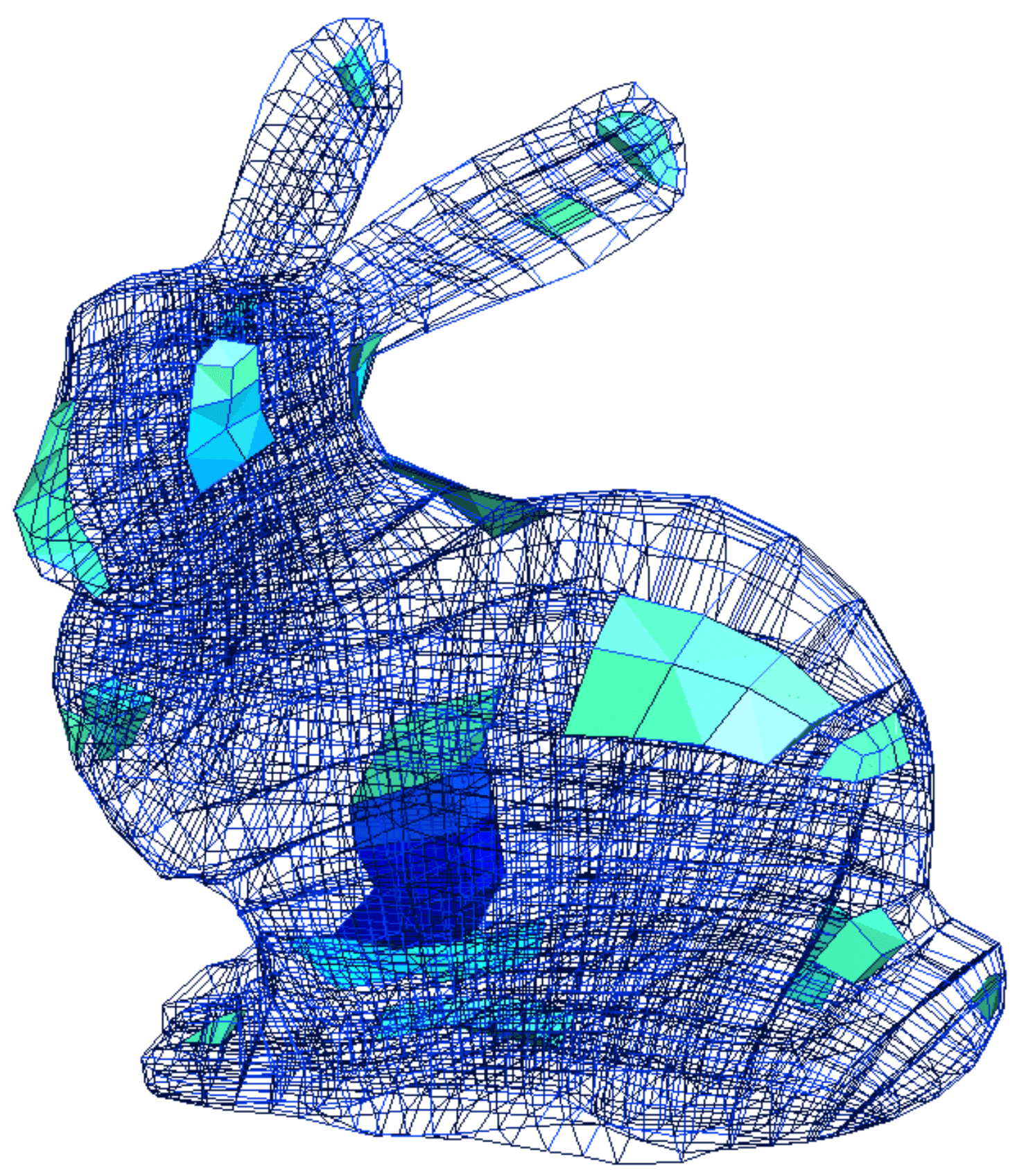}
\caption{invalid hexahedra in an example mesh}
\label{fig:bunny-blobs}
\end{figure}

For many cases where the mesh is not completely twisted, the number of invalid hexahedra is a fraction of the total number of hexahedra. For example, in Fig. \ref{fig:bunny-blobs}, where the filled-out hexahedra represent the invalid hexahedra. In this mesh, only 1.59\% of hexahedra are invalid. Also visible is how well these invalid hexahedra can be grouped up. Therefore, avoiding including all hexahedra in the untangling process seems reasonable.
To reduce the number of hexahedra included, it makes sense only to allow vertices close to the invalid hexahedra to move. By only allowing specific vertices to move, there are not as many elements that need to be included in the energy calculation. Because these elements remain the same and the untangling of the invalid hexahedra can therefore be done locally without the influence of these elements. This way, a lot of computational resources can be saved. There are two different strategies which can be applied.

\begin{figure}[H]
\centering
\includegraphics[scale=0.8]{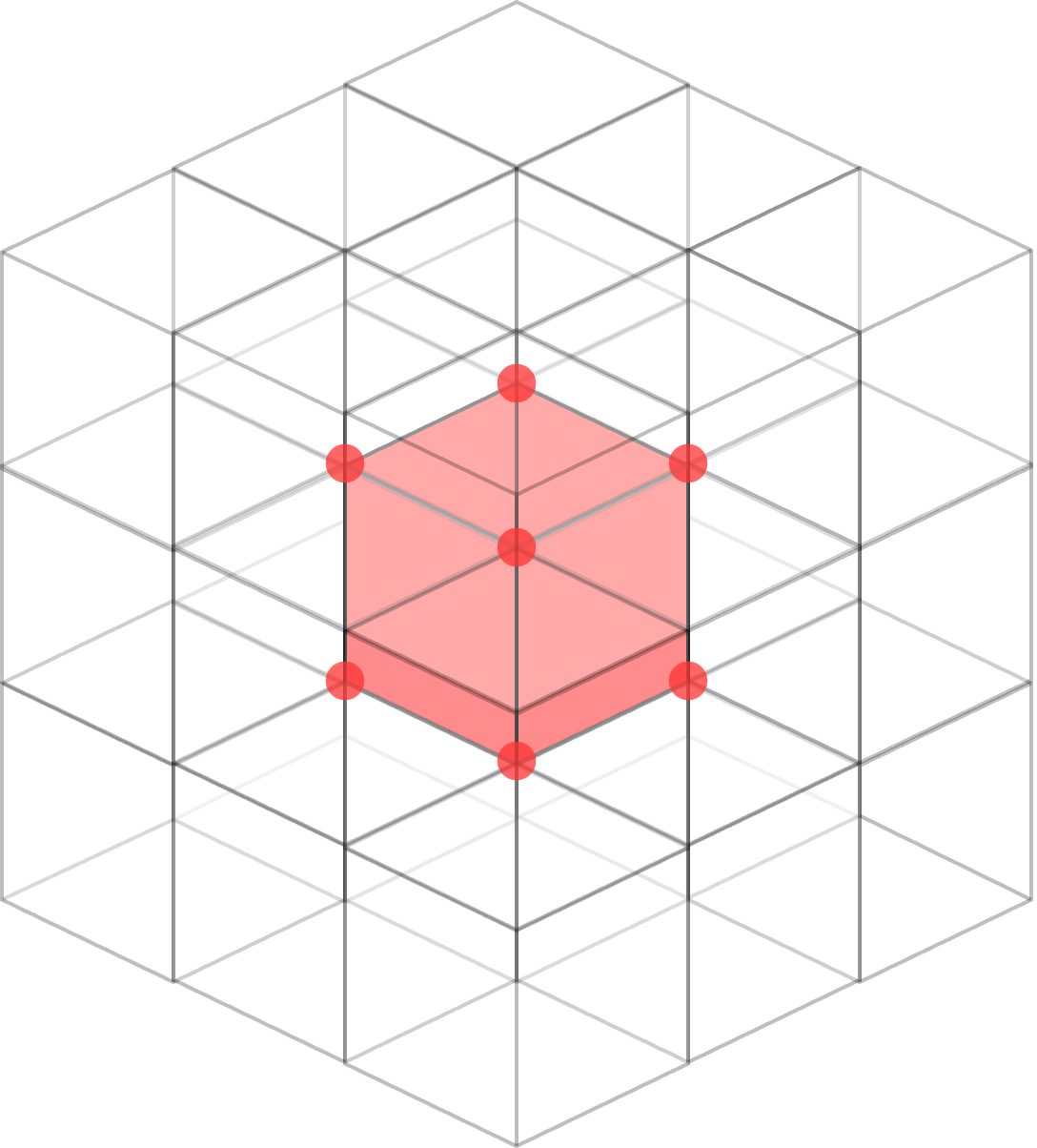}
\caption{an invalid hexahedron marked in red in the middle, being surrounded by its neighbours}
\label{fig:layer}
\end{figure}

First, the whole mesh is still optimised at every step, but the energy is calculated by only including the invalid hexahedra as well as all their neighbours. Neighbours here meaning all hexahedra which share a vertex with one invalid hexahedron. We call this strategy the "whole mesh blob strategy". The combination of an invalid hexahedron and its neighbours is what we call a blob. In Fig. \ref{fig:layer} such a blob is illustrated. The red hexahedron in the middle is the invalid one, the other black ones around it are its neighbours. But even in such a blob, not all vertices are free to move. Only the vertices which belong to an invalid hexahedron are allowed to be moved. Therefore the neighbours form a bubble inside which the vertices can be moved at will, but the borders stay in place. As only such few hexahedra are included, the amount of elements which need to be optimised decreases massively. In combination with fewer vertices being allowed to move, it speeds up the untangling process.

The second option is similar to the first, as the blobs are built in the same way. But this time, only one blob is created per iteration. The mesh is then untangled only using the energy from this one blob. Hence we will call it the "individual blob strategy". When this blob is untangled, the next blob is selected and untangled. So an additional loop has to be added looping over all blobs. In general, the whole mesh is optimised, but only the vertices from the invalid hexahedron in one blob are allowed to move. The energy also only consists of the invalid hexahedron as well as its neighbours. Also to note here is that each blob has its individual $\epsilon$ value. So resetting it to the starting value after a blob is untangled. This significantly speeds up one optimisation step. As such few elements and movable vertices are included. But a lot of optimisation steps have to be made. Another addition to this blob version is to build bigger blobs in case there are invalid hexahedra next to each other. This means that when collecting the neighbours of an invalid hexahedron, checking whether one of those neighbours is also invalid. If that is the case, the vertices of those two hexahedra are allowed to move, and the number of neighbours increases as the neighbours of both hexahedra have to be included. This can be continued until no invalid hexahedra are a neighbour anymore. This allows whole areas of invalid hexahedra to be untangled at one time, as it might otherwise be challenging to untangle each hexahedron by itself.

The tradeoff of using either one of these blob strategies is that fewer elements are involved in the untangling. Therefore it could be possible that not enough room can be created where a blob could be untangled. This problem would not arise if all elements were included, as no more room could be created. Room here is understood as space that can be created by moving vertices of valid elements, inside which invalid hexahedra can be untangled. As an invalid hexahedron might be too restricted by vertices that are not movable, and therefore no solution can be reached even if one existed. As blobs are built in a layered manner, with an invalid hexahedron in the middle and a layer of neighbours around it, it is relatively easy to create more room. In this basic case, there is one invalid hexahedron in the centre with its 8 vertices which are movable. As mentioned before, there is not much room to move in. Currently, just the neighbouring hexahedra are included in the energy contribution. So there is an invalid hexahedron in the centre, enclosed by its neighbouring hexahedra, which are mostly static as most of their vertices are not allowed to move. Meaning there is just 1 layer around the invalid hexahedron, as visualised in Fig. \ref{fig:layer}. Again, the hexahedron marked in red is the invalid one, and the ones in black are all the neighbours. Now this number of layers could be increased. So for 2 layers, all the vertices of the invalid hexahedron's neighbours become movable. At the same time, the neighbours to the neighbours become relevant as well. Most of their vertices are fixed, but they are now included in the energy calculation. This way, there are many more degrees of freedom, vertices which can be moved, and therefore more room to untangle the blob. This increment can be done if the untangling process is stuck in an effort to still reach a solution. An idea is to execute the increment when conditions \eqref{eq:elements-conditions} are fulfilled. For the first time these conditions are fulfilled, the amount of tetrahedra can be increased for all hexahedra, and from then on, the next times, the blob layer amount is increased.

To sum up the improvement of the usage of these blobs. When using the first blob strategy of untangling all blobs simultaneously, only 11.19\% of hexahedra in the mesh of Fig. \ref{fig:bunny-blobs} are needed to build all blobs necessary to untangle the mesh. This shows the potential of this strategy.

\newpage

\chapter{OpenFlipper Plugin Evaluation}
To see how well the algorithm performs, some test meshes are used. We show how our algorithm stacks up against the algorithm from \cite{sos} and how some factors influence the returned result.

Regarding the test cases, as mentioned in section \ref{chap:related-work}, we use the same 10 test cases as \cite{sos}.
Starting with the validity and amount of invalid elements of these meshes represented in Table \ref{tab:inv-hexa}.
\begin{table}[h]
\centering
\scalebox{0.8}{
\begin{tabular}{ |c|c|c|c|c| }
\hline
Case & Total \#Hexahedra & 
\multicolumn{3}{|c|}{Invalid \#Hexahedra} \\
\hline
 & & 58 Tets test (ours) & Bézier test \cite{validation} & SOS test \cite{sos} \\
\hline
Block & 2520 & 111 & 31 & 19 \\ 
Bunny & 2832 & 156 & 45 & 40 \\ 
Bust & 5258 & 334 & 30 & 30 \\
Cat & 2604 & 97 & 1 & 1 \\ 
Dolphin & 4788 & 96 & 2 & 2 \\ 
Elephant & 46525 & 2556 & 408 & 90 \\ 
Fertility & 21016 & 61 & 0 & 20 \\ 
Horse & 44145 & 1830 & 304 & 73 \\ 
Rockerarm & 1858 & 52 & 11 & 11 \\ 
Torus & 7380 & 133 & 121 & 121 \\ 
\hline
\end{tabular}}
\caption{the different test cases with the total number of hexahedra making up the mesh and then the number of invalid hexahedra according to the three different methods}
\label{tab:inv-hexa}
\end{table}
First to note is the big difference of invalid hexahedra between the two validity checking methods presented in this thesis, as mentioned in section \ref{sec:validity} already. But secondly, the difference between the Bézier test method and the SOS test method needs to be discussed. For half the cases, both methods classify the same amount of hexahedra as invalid. But for the other half, the differences are inconsistent. For 4 of the other 5 meshes, the sos paper method classifies fewer hexahedra as invalid than the improved method. But then there is the Fertility case where the Bézier algorithm does not detect any invalid elements while the sos paper classifies a few hexahedra as invalid. We suspect numerical inaccuracy to be the reason for this difference. Therefore, we side with the improved algorithm and leave out the Fertility case.

\section{Timings} \label{sec:test}

\paragraph{Setup}
The experimentation times are achieved with the following setup. The algorithm was implemented as a plugin for OpenFlipper \cite{openflipper} and was therefore written in C++. As our optimiser, we use the TruncatedNewtonPCG from CoMISo \cite{comiso}. For which the derivatives are computed by TinyAD \cite{tinyad}. Regarding the algorithm implemented, we use the global algorithm with all the improvements introduced in section \ref{sec:algorithm}, together with the "whole mesh blob strategy". The only exception is that the layer increase from section \ref{sec:blobs} is not implemented, as these test cases are solvable without it. Our implementation can be found on GitLab: \url{https://gitlab.inf.unibe.ch/cgg/code/plugin-hexoptimizer}

We present the results we got using the above setup and compare them to those achieved by the sos paper \cite{sos}.

\begin{table}[h]
\centering
\begin{tabular}{ |c|c|c|c|c| }
\hline
Case &
\multicolumn{2}{|c|}{Blob (Whole Mesh)} & 
\multicolumn{2}{|c|}{SOS Paper \cite{sos}} \\
\hline
 & Time & Iterations & Time & Iterations \\
\hline
Block & 2.72251 \textbf{s} & 2 & 2.29 \textbf{min} & 11 \\ 
Bunny & 39.7268 \textbf{s} & 52 & 4.05 \textbf{min} & 19 \\ 
Bust & 5.39762 \textbf{s} & 4 & 9.20 \textbf{min} & 49 \\
Cat & 9.9573 \textbf{s} & 37 & 1.41 \textbf{min} & 11 \\ 
Dolphin & 1.24474 \textbf{s} & 3 & 2.45 \textbf{min} & 2 \\ 
Elephant & 51.431 \textbf{s} & 6 & 24.85 \textbf{min} & 12 \\ 
Fertility & - & - & 9.83 \textbf{min} & 21 \\ 
Horse & 29.7245 \textbf{s} & 4 & 22.15 \textbf{min} & 12 \\ 
Rockerarm & 1.29553 \textbf{s} & 3 & 1.24 \textbf{min} & 7 \\ 
Torus & 3.7315 \textbf{s} & 1 & 12.13 \textbf{min} & 12 \\ 
\hline
\end{tabular}
\caption{timings from our method, the whole mesh blob strategy, plus the outer iterations needed to optimise the mesh compared to the sos results}
\label{tab:opt-times}
\end{table}

\begin{figure}[H]
\centering
\includegraphics[width=\textwidth]{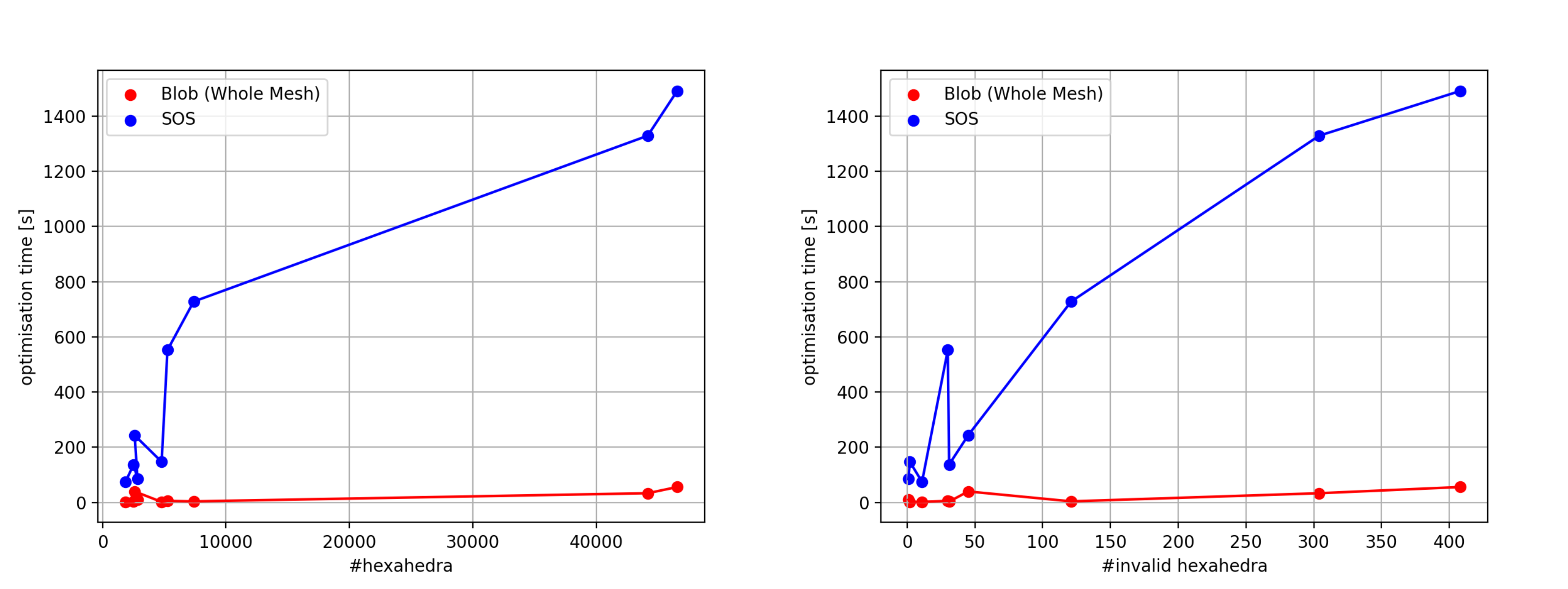}
\caption{performance of the algorithms: on the left is the optimisation time against the amount of hexahedra in a mesh, and on the right is the optimisation time against the amount of invalid hexahedra in a mesh}
\label{fig:algo-perf}
\end{figure}

Looking at the results from Table \ref{tab:opt-times}, the first thing noticed is that our algorithm is a lot faster than the sos paper algorithm. Second, the amount of outer iterations does not impact how fast an algorithm is, neither for our nor the sos algorithm. But also an important note is that the amount of outer iterations between the two algorithms cannot be compared as the algorithms do not work in the same way. This performance supremacy of our algorithm compared to the sos algorithm is also visualised in Fig. \ref{fig:algo-perf}. From these plots, it is pretty evident that against the first intuition, there does not seem to be a direct and only dependency on the mesh size or the amount of invalid hexahedra in a mesh. But there seem to be some more factors.

 As discussed above, there is not one simple factor determining how fast a mesh can be untangled. Naturally, the mesh size and the number of invalid hexahedra significantly affect the time needed, but those are not the only factors. One main factor, which is hard to quantify, is how badly tangled the situation is. As not every problem has the same difficulty level to untangle it. The cases where the boundary needs to be moved certainly take longer than cases where the boundary can stay locked. These kinds of factors certainly play a significant role in the optimisation time required as well. A good example where this is visible are the bunny and torus cases. The bunny mesh is much smaller than the torus mesh and has fewer invalid hexahedra. But even with those characteristics, it takes our method quite a bit longer to solve than the torus mesh.

\begin{figure}[H]
\centering
\includegraphics[width=\textwidth]{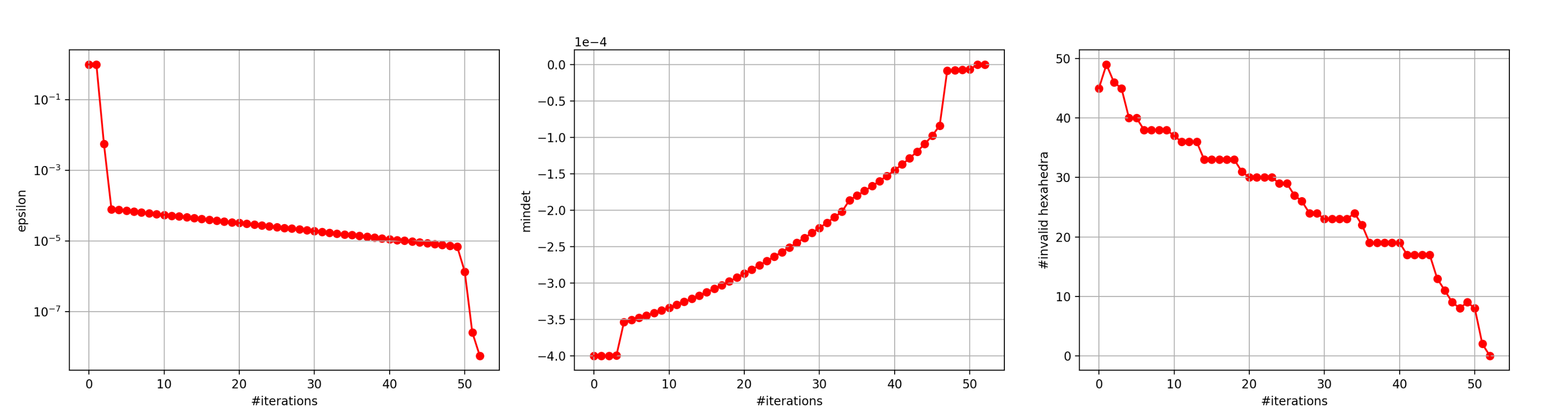}
\caption{the development of the three statistics during the untangling process for the bunny mesh, the epsilon value on the left (logarithmic scale), the minimum determinant value in the middle and the amount of invalid hexahedra on the right}
\label{fig:iter-stats}
\end{figure}

 During the untangling process, a lot of iterations are passed. During each of these iterations, some key values change. Their development is displayed here in Fig. \ref{fig:iter-stats} for the bunny mesh.
On the left is the epsilon value, which has a rather uncommon development. This is due to the improvement made in section \ref{sec:epsilon-update}. Otherwise, the descent of the epsilon value would not be this drastic. The minimum determinant in the middle plot increases nicely in every iteration, and the untangling process ends as soon as it hits zero. Lastly, the amount of invalid hexahedra on the right shows that at specific points, the mesh has to destroy some hexahedra in order to untangle other ones. In general, though, it decreases steadily, and the untangling process ends when zero invalid hexahedra are left.

\section{Boundary Relaxation} \label{sec:bnd-movement}
As discussed in section \ref{sec:movable-bnd}, the goal next to being able to untangle a mesh is to do it so that the surface changes as little as possible. As otherwise, the mesh might be untangled but end up looking nothing like it did in the beginning. This section closely examines how well our algorithm achieves this second goal.

\begin{table}[h]
\centering
\scalebox{0.8}{
\begin{tabular}{ |c|c|c|c| }
\hline
Case & \multicolumn{2}{|c|}{bnd movement (avg)} & bnd movement (max)\\ 
\hline
 & all & movable & \\
\hline
Block & 0.000000E+00 & 0.000000E+00 & 0.000000E+00 \\ 
Bunny & 1.685291E-04 & 1.351521E-03 & 3.200396E-02 \\ 
Bust & 9.987311E-05 & 3.242547E-02 & 9.426871E-02 \\
Cat & 2.486089E-05 & 2.272285E-03 & 9.502378E-03 \\ 
Dolphin & 1.978699E-06 & 2.777374E-04 & 7.348913E-04 \\ 
Elephant & 4.216888E-05 & 1.842480E-03 & 6.857785E-03 \\ 
Fertility & - & - & - \\ 
Horse & 8.254699E-05 & 1.657444E-03 & 6.297005E-03 \\ 
Rockerarm & 4.175861E-05 & 1.585756E-03 & 4.642058E-03 \\ 
Torus & 0.000000E+00 & 0.000000E+00 & 0.000000E+00 \\ 
\hline
\end{tabular}}
\caption{the distance the boundary vertices have moved compared to the input mesh, first the average movement of all boundary vertices, then the average movement of only the movable boundary vertices and third the movement of the vertex which has moved the most}
\label{tab:bnd-movement}
\end{table}

Table \ref{tab:bnd-movement} shows how much all boundary vertices move. This distance is computed by:
$$d = ||\boldsymbol r_{end} - \boldsymbol r_{start}||_2 = \sqrt{(x_{end} - x_{start})^2 + (y_{end} - y_{start})^2 + (z_{end} - z_{start})^2}$$
Where x,y, and z are always the three components of the vertex coordinates at the start, in the original mesh, and at the end, after the optimisation.

Looking at these results, the average boundary vertex moves a small but noticeable distance. This distance stays at zero for meshes where it is not necessary to move the boundary vertices at all, which is precisely what is expected. Looking at the average distance of only the movable boundary vertices, the distance naturally gets bigger and, in some cases, even gets into the same order as the maximal movement. In general, this value is quite a bit bigger than we would wish for, as the mesh surface is deformed a lot, sometimes even enough to be visible. Therefore, the goal of being able to untangle a mesh while not deforming the original surface too much is not satisfyingly achieved.

In order to put these boundary vertex movements into perspective, they have to be scaled somehow. Because the vertices move a lot more if the mesh is saved on a bigger scale, as the distances between vertices are big, compared to when it is saved on a small scale. Note that the goal of this scaling is not to increase or decrease the distance the vertices move. But simply to make these values comparable. Therefore a new scaler $\overline{l}$ is introduced for every boundary vertex. This scaler is computed by:
$$\overline{l} = \frac{\sum_{i=1}^n ||\boldsymbol r_i - \boldsymbol r_0||_2}{n}$$
Where $\boldsymbol r_0$ is the vertex the scaler is computed for, $\boldsymbol r_i$ are the neighbouring vertices of $\boldsymbol r_0$, which are also boundary vertices and n is the total amount of the boundary vertex neighbours. This scaler is, therefore, an average distance between the vertex and its boundary neighbours. To note here is that this scaler is computed on the input mesh before any untangling takes place. After the untangling is done, the boundary movement distance is calculated as before. But this time, it is scaled by:
$$d_{scaled} = \frac{d}{\overline{l}}$$
The results of this scaled distance are visible in Table \ref{tab:scaled-bnd-movement}. These results show once more that the boundary is moved a lot more than we would like. The biggest movements are done on the horse mesh. This is also visible in Fig. \ref{fig:horse-comparison}.

\begin{table}[h]
\centering
\begin{tabular}{ |c|c|c|c| }
\hline
Case & \multicolumn{2}{|c|}{bnd movement scaled (avg)} & bnd movement scaled (max)\\ 
\hline
 & all & movable & \\
\hline
Block & 0.000000E+00 & 0.000000E+00 & 0.000000E+00 \\ 
Bunny & 4.164403E-03 & 3.339648E-02 & 3.831065E-01 \\ 
Bust & 5.653721E-05 & 1.835575E-02 & 5.047078E-02 \\
Cat & 1.857086E-04 & 1.697377E-02 & 7.908092E-02 \\ 
Dolphin & 1.171649E-05 & 1.644569E-03 & 4.158615E-03 \\ 
Elephant & 4.791739E-03 & 2.093649E-01 & 5.898279E-01 \\ 
Fertility & - & - & - \\ 
Horse & 8.840065E-03 & 1.774979E-01 & 7.445004E-01 \\ 
Rockerarm & 1.136450E-03 & 4.315598E-02 & 1.293091E-01 \\ 
Torus & 0.000000E+00 & 0.000000E+00 & 0.000000E+00 \\ 
\hline
\end{tabular}
\caption{the scaled distance the boundary vertices have moved compared to the input mesh, first the scaled average movement of either all boundary vertices or only the movable boundary vertices and second the scaled movement of the vertex which has moved the most}
\label{tab:scaled-bnd-movement}
\end{table}

Another test to see the movement of boundary vertices can be taken from the "Edge-Cone Rectification" paper\cite{edge-cone}. They prepared some stress test cases where a valid mesh is broken entirely by moving all vertices except boundary vertices. The goal for such a mesh would be to return to its valid form without having to move any boundary vertices.
\begin{figure}[H]
\centering
\includegraphics[scale=0.7]{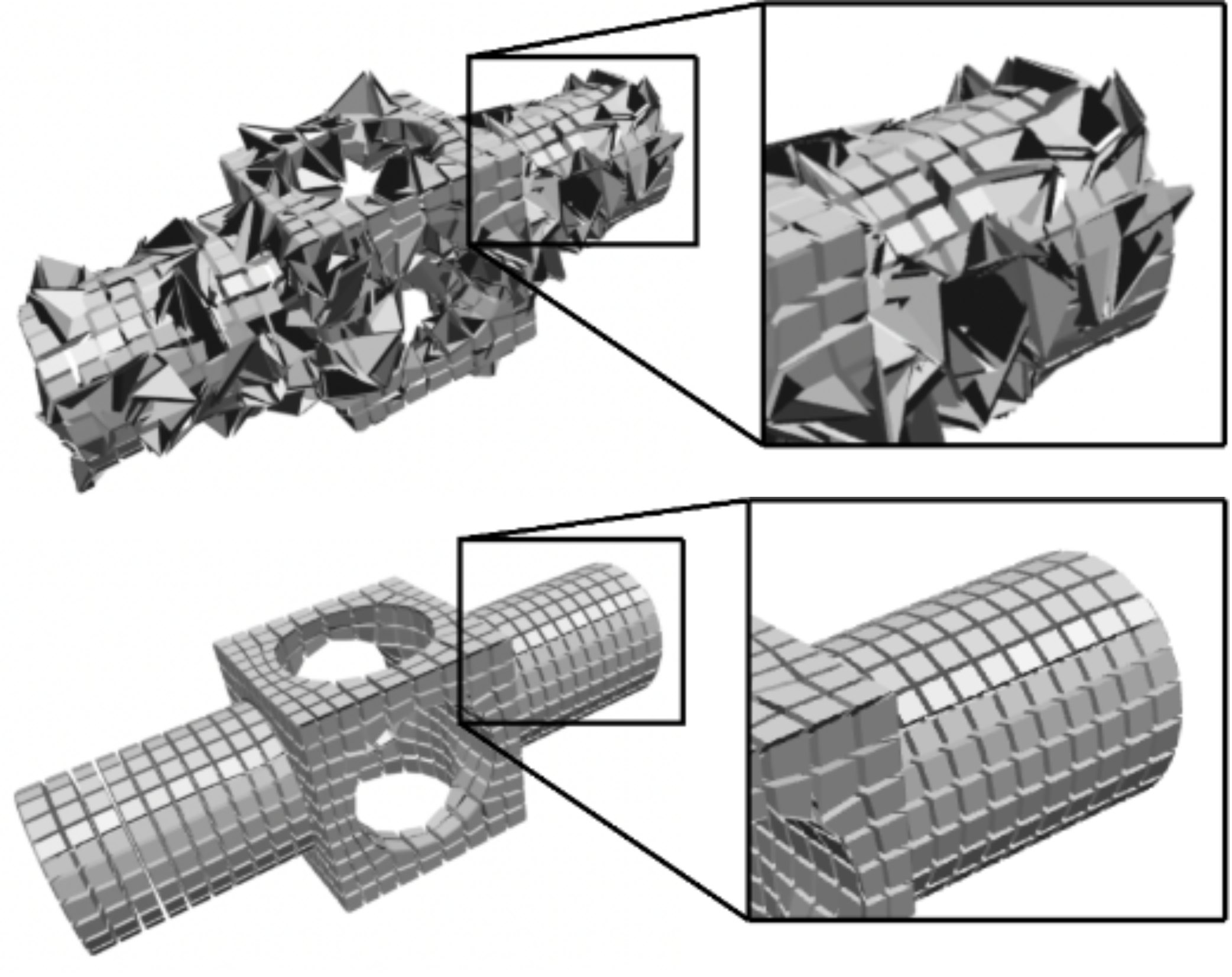}
\caption{the block stress case, before and after being untangled}
\label{fig:stress-test}
\end{figure}
\noindent
As shown in Fig. \ref{fig:stress-test}, the mesh is untangled without being deformed, as the movement of the boundary vertices stays at zero. Therefore the goal of finding a solution without moving the boundary vertices over one where the boundary vertices are moved is achieved. This is what we hope for, as otherwise, the algorithm does not get penalised enough for moving boundary vertices.

\subsection{Penalty Term Factor} \label{sec:influence-penalty-factor}
\begin{figure}[H]
\centering
\includegraphics[scale=0.475]{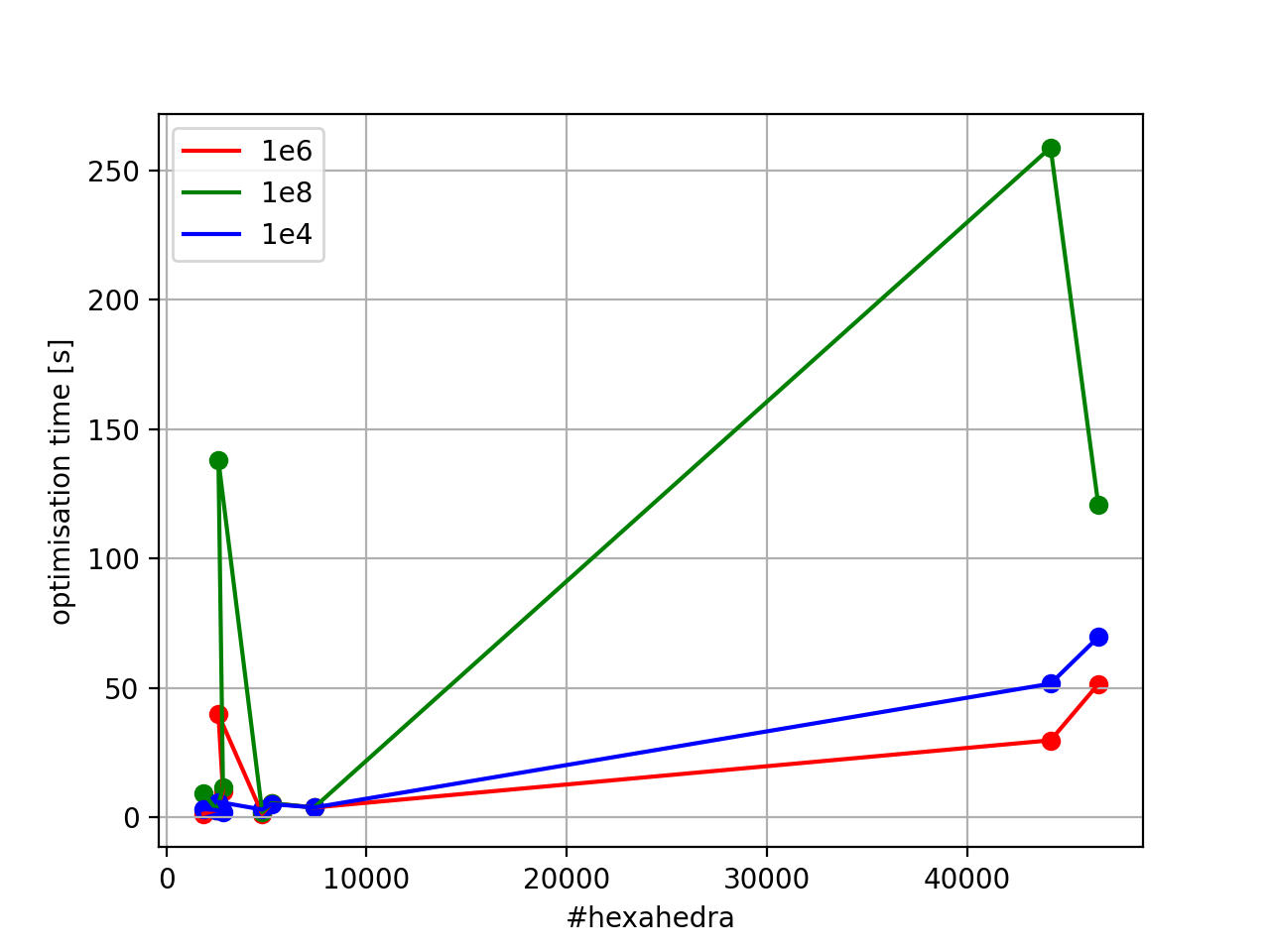}
\caption{the time it takes to untangle the meshes with different values for the penalty factor}
\label{fig:factr-times}
\end{figure}

This factor plays a significant role in the surface deformation and, therefore, the deformation of the whole mesh. The factor of $10^6$ we suggested in section \ref{sec:movable-bnd} was empirically determined. It is the factor which can untangle the 10 test cases and minimise the deformation of their boundary while the time to untangle stays reasonable. As the time needed to untangle a mesh changes significantly with other values. As visualised in Fig. \ref{fig:factr-times}, where the times for a lower $10^4$ and a higher $10^8$ factor are also shown. Interestingly, a lower factor does not necessarily result in a faster untangling process on bigger meshes. This again proves the algorithm's unpredictability and shows it is contingent on all different kinds of factors.

\begin{figure}[H]
\centering
\includegraphics[width=\textwidth]{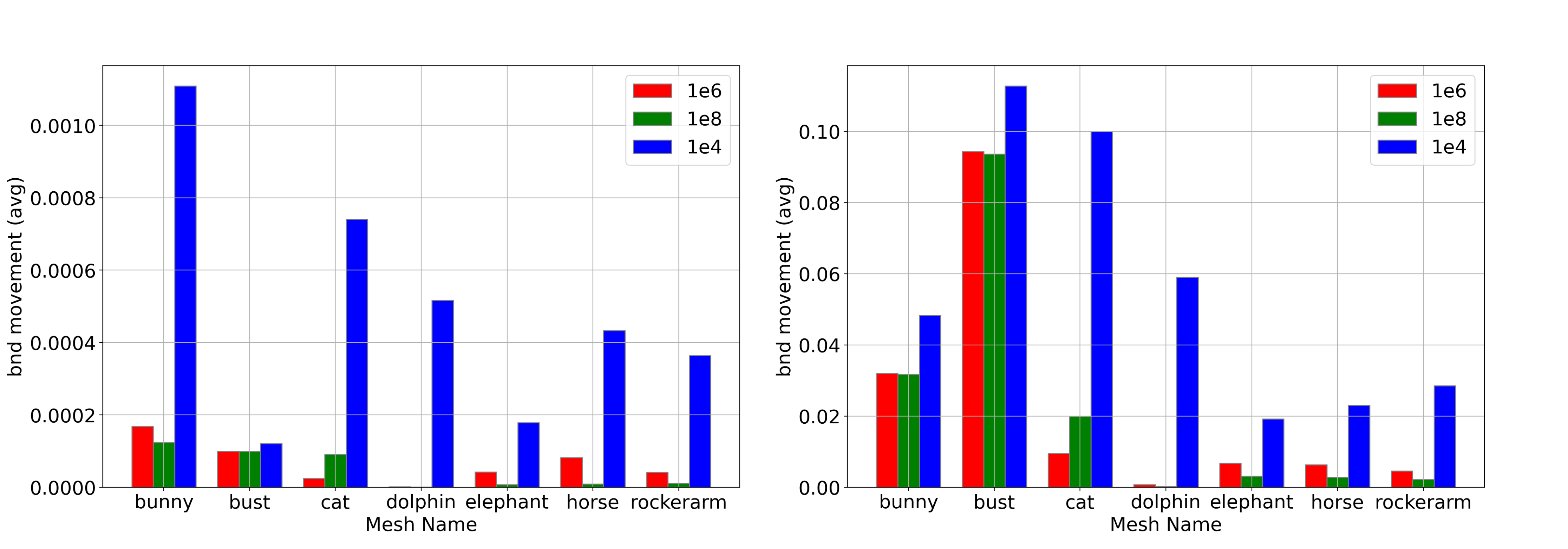}
\caption{the average boundary movement on the left and the maximal boundary movement on the right}
\label{fig:factr-acc}
\end{figure}

\begin{figure}[H]
\centering
\includegraphics[scale=0.7]{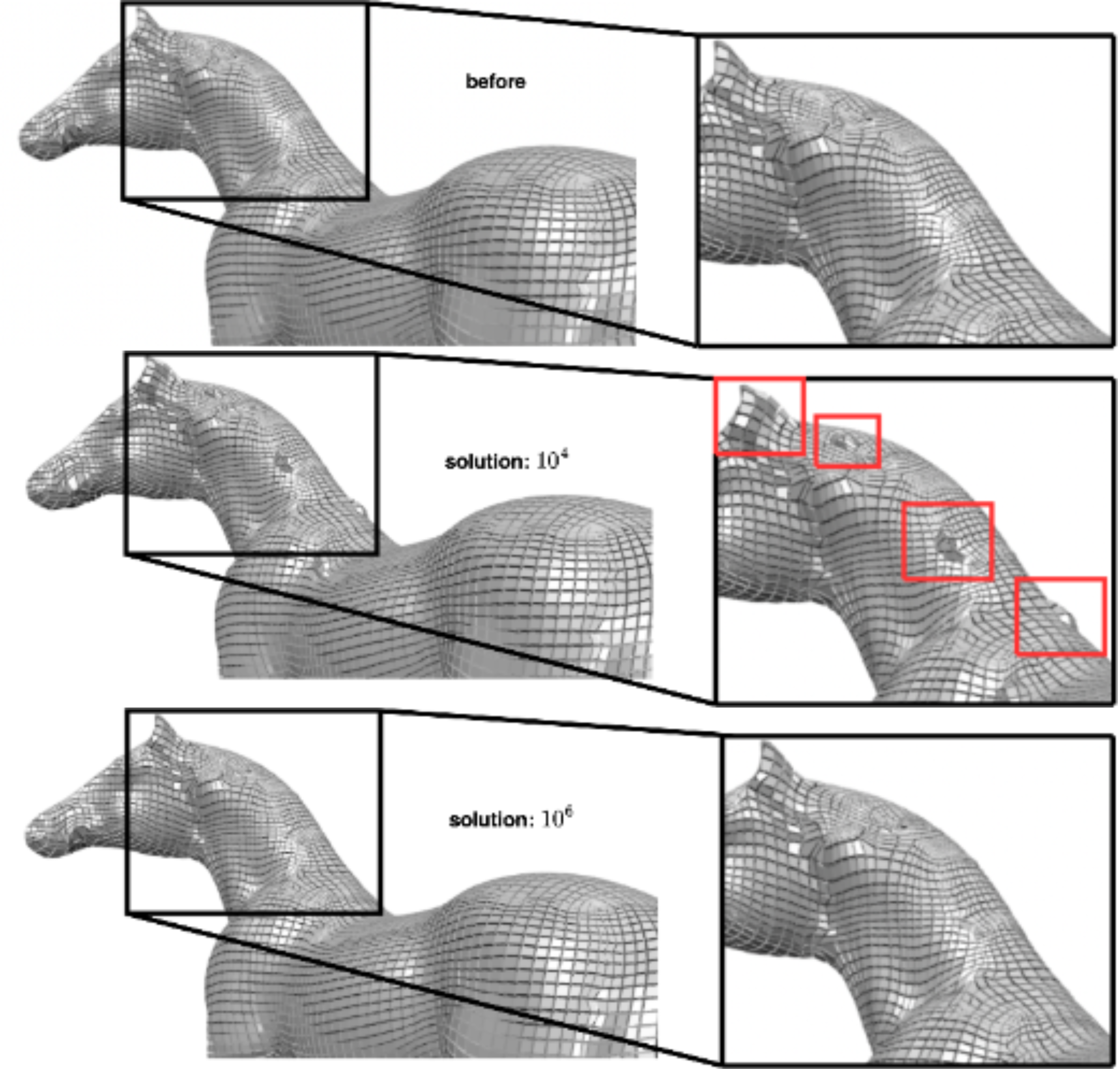}
\caption{the horse mesh before the untangling up top, the untangled solution with a factor $10^4$ in the middle and the untangled solution for a factor $10^6$ on the bottom}
\label{fig:horse-comparison}
\end{figure}

Now while the timings are interesting, they are not one of the main goals. More critical is the deformation of the mesh. Therefore in Fig. \ref{fig:factr-acc}, it is visualised how much the boundary vertices are moved. Again for these same three factors, $10^6$, $10^8$ and $10^4$. On the left is the average boundary vertex movement and on the right is the maximum boundary vertex movement. Important to note here is that the two meshes, block and torus, which can be solved without moving boundary vertices, have been left out of the figures.
As shown in Fig. \ref{fig:factr-acc}, the factors $10^6$ and $10^8$ always deliver very similar results. But with the above timing drawback, the penalty factor $10^6$ seems the most reasonable as it is the best balance between speed and accuracy. It is also clearly visible in Fig. \ref{fig:factr-acc} how the mesh deformation increases significantly with a lower penalty factor. This drop in quality can even be visualised, as seen in Fig. \ref{fig:horse-comparison}.
In this figure, the solution for a factor of $10^4$ shows some bubbles coming out of the horse's neck, as well as the ear being deformed quite a lot. While the solution for a factor of $10^6$ shows no such thing. This indicates that a penalty factor which is too low can prefer solutions where boundary vertices are moved instead of trying a little harder and finding a solution where these boundary vertices do not need to be moved.

Therefore the penalty factor $10^6$ seems to be the best option. But it has to be remembered that it is the best for these 10 test cases and might be different for other meshes.

\section{Influence of Variables} \label{sec:influence-variables}
As seen during the construction of the algorithm in section \ref{sec:algorithm}, many optimisations can be made to speed up the algorithm. But there are a lot of variables which affect the algorithm's output as well as how long it takes to get this output. It would be out of scope to try and analyse all these variables in detail. That is why only a few select ones are discussed here, as they also allow for further experimentation.

\subsection{Inner Iterations}
\begin{figure}[H]
\centering
\includegraphics[scale=0.5]{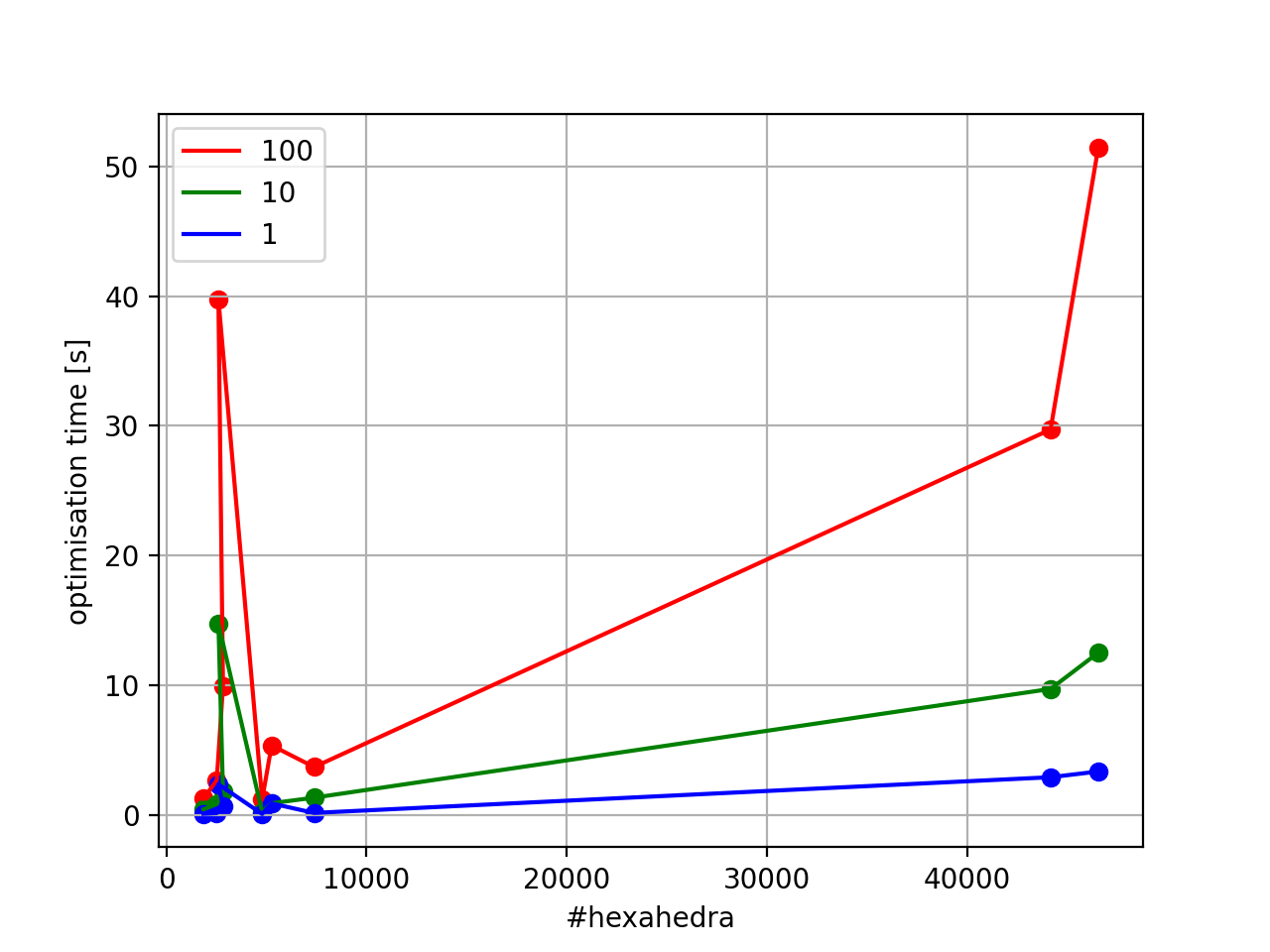}
\caption{the time it takes to untangle the meshes with different values for the amount of inner iterations}
\label{fig:inner-times}
\end{figure}

The TruncatedNewtonPCG optimiser used for the untangling has an inner iterations parameter. To understand what exactly these inner iterations stand for, a closer look at the optimiser would be necessary, which would be out of scope for this thesis. Therefore the basic understanding of a small number of inner iterations solving a problem less accurately than a big number of inner iterations should suffice. Throughout the development of the algorithm, we realised that a small number of inner iterations, even though not solving the problem very exactly, converged incredibly fast for these 10 test cases. As is visible in Fig. \ref{fig:inner-times} where the times to untangle the meshes are plotted for inner iterations values of 100, 10 and 1. The original timings in Table \ref{tab:opt-times} where done with 100 inner iterations. This is because 100 inner iterations allow for some more complex meshes to be solved, like the dragon mesh from \cite{edge-cone}, and therefore proofs to be a more general and robust value.
As mentioned before, a less accurate solution should be reached if fewer inner iterations are used. To check whether this is the case for this problem, the boundary vertex movement from section \ref{sec:bnd-movement} is used again. We are using the general non-scaled average movement of all boundary vertices as this serves the purpose of comparing the same meshes with a different factor. So only the average boundary vertex movement is shown in Table \ref{tab:bnd-movement-inner} for the three inner iterations values.

\begin{table}[h]
\centering
\begin{tabular}{ |c|c|c|c| }
\hline
Case & 100 & 10 & 1 \\
\hline
Block & 0.000000E+00 & 0.000000E+00 & 2.463277E-07 \\ 
Bunny & 1.685291E-04 & 1.682539E-04 & 1.566630E-04 \\ 
Bust & 9.987311E-05 & 9.987036E-05 & 3.502917E-04 \\
Cat & 2.486089E-05 & 2.486089E-05 & 2.480772E-05 \\ 
Dolphin & 1.978699E-06 & 1.978698E-06 & 1.975698E-06 \\ 
Elephant & 4.216888E-05 & 4.215501E-05 & 3.957151E-05 \\ 
Fertility & - & - & - \\ 
Horse & 8.254699E-05 & 8.254698E-05 & 7.129522E-05 \\ 
Rockerarm & 4.175861E-05 & 4.175860E-05 & 3.823587E-05 \\ 
Torus & 0.000000E+00 & 0.000000E+00 & 0.000000E+00 \\ 
\hline
\end{tabular}
\caption{the non-scaled average distance all boundary vertices have moved during the untangling process when using the three different inner iterations values}
\label{tab:bnd-movement-inner}
\end{table}

When looking at these results, the first thing noticed is that the solutions reached for 10 and 100 inner iterations seem identical. Which would suggest 10 as a better value since the timings with 10 are significantly faster. When only using 1 inner iteration, on the other hand, the quality of the solution drops. Firstly, when only using 1 inner iteration, there are boundary vertices moved in the untangling of the block mesh. Which should not be necessary, as the other values show. Additionally, on average, the boundary vertices are moved more to untangle the meshes than when using more inner iterations. So 1 does not seem like the best option. Therefore 10 would look like the best option. Though it has to be remarked that 10 is the best option for the 10 test cases we analysed, and it cannot be generalised. But as mentioned before, 100 allows more complex meshes to be solved. For the cases examined here, this does not matter, as they are pretty easy to solve. But on more complex meshes, 10 inner iterations might not be enough to reach a solution, whereas 100 might. This is why an inner iterations value of 100 is chosen.

\subsection{Amount of Tetrahedra/Hexahedron} \label{sec:amount-tets}
At the beginning of this thesis, the expectation was that the amount of elements chosen for the optimisation would play a more prominent role. We did not expect to be able to solve most meshes while only using the necessary 8 corner tetrahedra. For this reason, some experiments were planned to figure out which amount of elements would be optimal for the optimisation. As the fewest amount of elements always end up in the fastest solution, this feels like an unnecessary step.

Though, as mentioned in section \ref{sec:amount-elements}, there might still be some application, as the rise of tetrahedra used from 8 to 58 might be too much. Some experiments could be run to see whether there is a number of tetrahedra lower than 58, allowing for a faster untangling. Some ideas are talked about in the future work paragraph at the end.

As the amount of elements used during the optimisation affects the amount of time it takes to untangle a mesh heavily, this might still be a worthwhile step. But in the end, it is considered out of scope for this thesis.

\newpage

\chapter{Conclusion}
Our approach is based on the results of mainly two papers. Combining each of their results allowed for an algorithm to be constructed which could untangle hexahedral meshes. The base algorithm introduced at the beginning of section \ref{sec:algorithm} met our goal, but not very efficiently and with many limitations. Therefore we introduced multiple improvements to make this approach more viable. This final algorithm, described in the "Setup" of section \ref{sec:test}, performs better than the base algorithm. It also performs better than algorithms with the same goal but different approaches, like the algorithm from the "SOS" paper \cite{sos}.

This thesis aimed to find a way to untangle hexahedral meshes in an efficient and non-deforming way. There are two conclusions to be drawn. Based on the results, the efficiency part is achieved. The non-deforming part, on the other hand, is not achieved quite as well, as the surface is altered significantly. In turn, these two goals were partially reached with definite room for improvement.

The first goal of an efficient untangling algorithm was achieved. In comparison to other algorithms, our algorithm performs very well. This improvement is due to the different approach, but also due to the multitude of improvements we introduced. Therefore this is considered to be a success.

Regarding the second goal, as this algorithm is supposed to be used as a step in the automatic generation of hexahedral meshes, it should be usable for all invalid meshes. But as discussed, if the boundary of the original mesh is not valid, the algorithm moves the boundary vertices too far to untangle the mesh. Sadly, the surface is invalid for most incoming meshes, as this is where the mesh-generating algorithm has the biggest problems generating a valid solution itself. Therefore this algorithm is not able to perform the task it is assigned to up to the standard which would be expected.

In general, this thesis presents another way to untangle hexahedral meshes. It is not the final solution to the problem it set out to solve, but it might prove to be a stepping stone for future work. Either as the base for another algorithm or simply showing that this approach is not good enough and resources should be invested elsewhere.

\paragraph{Future Work}
For future work, it seems logical to improve upon the problem of surface deformation, as this appears to be the most significant limitation of the algorithm. Some further ideas which could be explored here to keep the surface as similar as possible to the original surface. Firstly when sticking with the approach presented here, some of the vertices could possibly be locked back up again in case they do not move a lot. We did not feel the need to take this step, but it could be worth exploring. Another idea would be to reduce the number of boundary vertices which are allowed to move even further. As theoretically, only the boundary vertices connected to an invalid surface quadrangle and those connected to an invalid hexahedron with more than 4 boundary vertices are definitely needed in order to move to untangle the mesh. We explored this approach for a bit, but it prevented us from untangling meshes which were possible with the approach presented above. In order to not extend the timeframe of this thesis much longer, this approach was then dropped. Building upon one of these approaches, it is also possible to restrict the direction in which the boundary vertices are allowed to move. As the movement of a boundary vertex in the plane of the surface is not as detrimental as the movement away from this plane. This can be realised through another penalty added depending on the direction the vertex moves in. Penalising the movement away from the plane more.

Another topic worth possible improvement would be the amount of tetrahedra used per hexahedron. One interesting approach could be adding all the tetrahedra with an invalid Jacobian determinant on top of the 8 corner tetrahedra. Another possibility would be to run experiments on either whole meshes or individual hexahedra. During which, the amount of used tetrahedra changes and the times are recorded. This way, some possible combinations of tetrahedra which are more efficient than others could be found. To find these combinations, it could even be possible to use some help from a machine learning algorithm.

There are still some possible improvements left to explore. But it is also important to state here that this algorithm can untangle all meshes we expected it to untangle and do so very efficiently.

\end{document}